\documentclass[twocolumn]{aastex63}

\usepackage{graphicx}
\usepackage{array}
\usepackage{amssymb}
\usepackage{natbib}
\usepackage{float}
\usepackage{color}
\usepackage[utf8]{inputenc}  
\usepackage{amsmath}  
\usepackage{comment}
\usepackage{footnote}
\usepackage{longtable, tabu}
\usepackage{url}
\usepackage{pbox}
\usepackage[caption2]{ccaption}
\usepackage[figuresright]{rotating}
\usepackage{chngcntr}

\renewcommand*{\thefootnote}{\fnsymbol{footnote}}

\defcitealias{2020ApJS..249...17R}{R20}

\begin{document}

\title{High-redshift Narrow-line Seyfert 1 Galaxies: A Candidate Sample}

\author{Suvendu Rakshit}
\affiliation{Finnish Centre for Astronomy with ESO (FINCA), University of Turku, Quantum, Vesilinnantie 5, 20014 University of Turku, Finland}
\affiliation{Aryabhatta Research Institute of Observational Sciences, Manora Peak, Nainital 263002, India}
\author{C. S. Stalin}
\affiliation{Indian Institute of Astrophysics, Block II, Koramangala, Bangalore-560034, India}
\author{Jari Kotilainen}
\affiliation{Finnish Centre for Astronomy with ESO (FINCA), University of Turku, Quantum, Vesilinnantie 5, 20014 University of Turku, Finland}
\affiliation{Tuorla Observatory, Department of Physics and Astronomy, FI-20014 University of Turku, Finland}
\author{Jaejin Shin}
\affiliation{Department of Astronomy and Atmospheric Sciences, Kyungpook National University, Daegu 41566, Republic of Korea}

\correspondingauthor{Suvendu Rakshit}
\email{suvenduat@gmail.com}
\shorttitle{high-z NLS1}
\shortauthors{Rakshit et al.}

\begin{abstract}

The study of narrow-line Seyfert 1 galaxies (NLS1s) is now mostly limited to 
low redshift ($z<0.8$) because their definition requires the presence 
of the H$\beta$ emission line, which is redshifted out of the spectral coverage of 
major ground-based spectroscopic surveys at $z>0.8$. We studied the 
correlation between the properties of H$\beta$ and Mg II lines of a large
sample of SDSS DR14 quasars to find high-$z$ NLS1 candidates. 
Based on the strong correlation of 
$\mathrm{FWHM(MgII)=(0.880\pm 0.005) \times FWHM(H\beta)+ (0.438\pm0.018)}$, 
we present a sample of high-$z$ NLS1 candidates having FWHM of Mg II $<$ 
2000 km s$^{-1}$. The high-$z$ sample contains 2684 NLS1s with redshift $z=0.8-2.5$ 
with a median logarithmic bolometric luminosity of 
$46.16\pm0.42$ erg s$^{-1}$, logarithmic black hole mass of $8.01\pm0.35 M_{\odot}$, and logarithmic Eddington ratio of $0.02\pm0.27$. The fraction of radio-detected high-$z$ 
NLS1s is similar to that of the low-$z$ NLS1s and SDSS DR14 quasars at a similar 
redshift range, and their radio luminosity is found to be 
strongly correlated with their black hole mass.  

\end{abstract}


\section{Introduction}

Narrow-line Seyfert 1 galaxies (NLS1s) are a special class of active galactic 
nuclei (AGNs), which are characterized by narrow permitted emission lines with 
the  full width at half maximum (FWHM) of the permitted  H$\beta<2000$ km s$^{-1}$ and 
flux ratio [O III]/H$\beta <3$ \citep{1983ApJ...273..478O,1989ApJ...342..224G}. 
They show stronger Fe II emission, higher amplitude rapid X-ray variability,
higher soft X-ray excess, and steeper soft and hard X-ray spectra compared to the broad line 
Seyfert 1 galaxies \citep[BLS1s; e.g.,][]{1994MNRAS.268..405N,1996A&A...305...53B,1999ApJS..125..317L}.
They are widely believed to have low black hole masses ($M_{\mathrm{BH}}$ 
$<$ 10$^8$ M$_{\odot}$) and  high accretion rates greater than  0.1 $L_{\mathrm{Edd}}$, 
where $L_{\mathrm{Edd}}$ is the Eddington luminosity defined as $L_{\mathrm{Edd}} = 1.3 \times
10^{38} \left(\frac{M_{\mathrm{BH}}}{M_{\odot}}\right)$ erg s$^{-1}$
compared to BLS1s \citep[e.g.,][]{2004ApJ...606L..41G,2006ApJS..166..128Z,2012AJ....143...83X,2017ApJS..229...39R}.  It has been found that the optical and infrared 
variability of NLS1s is lower compared to BLS1s primarily due to the higher 
Eddington ratio in the former \citep[see][]{2017ApJ...842...96R,2019MNRAS.483.2362R}. The low $M_{\mathrm{BH}}$ values and high accretion rates in NLS1s indicate that 
they are young and growing AGN and like radio-quiet AGN may not be able to 
produce relativistic jets \citep{2000MNRAS.314L..17M,2000NewAR..44..455G}.  
The low $M_{\mathrm{BH}}$ in NLS1s could be due to the projection effects \citep{2017ApJS..229...39R,2008MNRAS.386L..15D}. Spectropolarimetric observations \citep{2016MNRAS.458L..69B}, as well as modeling of accretion disk spectra \citep{2013MNRAS.431..210C}, indicate that the $M_{\mathrm{BH}}$ values of NLS1s available in the literature are an underestimation. Recently, \cite{2019ApJ...881L..24V} modeled a large sample of radio-loud NLS1s along with a control sample of radio-quiet NLS1s and BLS1s, and found that NLS1s have $M_{\mathrm{BH}}$ and Eddington ratio similar to BLS1s.

Only 7\% NLS1s are detected in radio surveys \citep{2006AJ....132..531K,2017ApJS..229...39R,2018MNRAS.480.1796S}. This is lower than the fraction of 
$\sim$15\% radio-loud sources found in normal AGN population \citep{1989AJ.....98.1195K}. 
About two dozen NLS1s show extended radio emission larger than 
20 kpc \citep[see][and the references therein]{2018ApJ...869..173R}. 
A minority of radio-loud NLS1s are also detected in the GeV $\gamma$-ray
band by the {\it Fermi} Gamma Ray Space Telescope unambiguously arguing
for the presence of relativistic jets in them. As of today about a 
dozen NLS1s are known to be emitters of $\gamma$-rays  
\citep[see][and the references therein]{2018ApJ...853L...2P}. Broad band
spectral energy distribution modeling of these $\gamma$-ray emitting
NLS1s show that they have the typical two hump structure with the high energy
emission due to inverse Compton scattering of seed photons external to the jets
of these sources \citep{2013ApJ...768...52P,2019ApJ...872..169P}. The broadband
SEDs of $\gamma$-ray emitting NLSy1s are similar to the flat spectrum radio 
quasar (FSRQ) category of AGN. Moreover, NLS1s are predominantly hosted by disk-like 
galaxies  \citep{2018A&A...619A..69J,2020MNRAS.492.1450O} while FSRQs are hosted 
in elliptical galaxies \citep{2007ApJ...658..815S}. If it were to be confirmed
that the host of $\gamma$-ray emitting NLS1s are indeed spirals, then we could 
conclude that relativistic jets are invariably launched by spiral as well
as elliptical hosts.

Our current knowledge of the general physical properties of  NLS1s is 
based on the  number of sources that we know up to $z$ = 0.8 \citep{2002AJ....124.3042W,2006ApJS..166..128Z,2017ApJ...842...96R,2018A&A...615A.167C}. Although only a couple of NLS1s are known beyond $z$ = 0.8 \citep{2015MNRAS.454L..16Y,2019MNRAS.487L..40Y}, it is possible that there are $\gamma$-ray emitting NLS1s beyond $z$ = 1. Since
many of the properties of low-$z$ NLS1s are similar to FSRQs and as FSRQs are
known up to large redshifts, it is natural to expect high-$z$ $\gamma$-ray
emitting NLS1s.  On the detection of such high-$z$ $\gamma$-ray emitting NLS1s, 
one can also check if the hosts of high-$z$ NLS1s are predominantly hosted by 
disk galaxies similar to their low-$z$ counterparts.  However, detecting 
$\gamma$-ray emitting NLS1s beyond $z = 0.8$ is hampered by the non-existent of 
optically known NLS1s with $z > 0.8$. One approach to find more $\gamma$-ray emitting 
NLS1 galaxies at $z > 0.8$ is first to arrive at a catalog of high-$z$ NLS1s 
in the optical and then look for their counterparts in the 10 years of data in 
{\it Fermi}. Our motivation in this work is therefore to arrive at a new catalog 
of high-$z$ NLS1s.  However, to find NLS1 at $z>0.8$, an alternative criterion is 
needed wherein another strong permitted emission line could serve as a proxy 
for H$\beta$. Recently, \citet[][hereafter \citetalias{2020ApJS..249...17R}]{2020ApJS..249...17R} found that the width of the Mg II 
line (available along with H$\beta$ line in the SDSS 
spectra in the redshift range of $z=0.35-0.8$) shows a strong correlation 
with  the line width of H$\beta$. Thus, in this paper, we investigated in detail 
the correlation between Mg II and H$\beta$ line properties of SDSS DR14 
quasars to arrive at a sample of high-$z$ NLS1s in the redshift range of $0.8-2.5$ 
based on Mg II. The structure of the paper is as follows. Section 
\ref{sec:sample_data} describes the sample and data,  Section \ref{sec:line_prop} 
investigates the properties of Mg II and H$\beta$ lines. 
Section \ref{sec:high-z-nls1} 
describes the criteria for high-$z$ NLS1 candidate selection and their multi-band 
properties are given in Section \ref{sec:prop}. We provide a summary in Section \ref{sec:summary}.

\section{Sample and data}\label{sec:sample_data}
To find high-$z$ NLS1s, we used the recently compiled catalog of the properties 
of about 500,000 quasars from  SDSS DR14 \citep{2018A&A...613A..51P} by 
\citetalias{2020ApJS..249...17R}. This catalog includes spectral properties such as 
emission line width and luminosity, and continuum properties such as  black hole mass, 
Eddington ratio, etc. This was done via a careful and systematic spectral modeling 
of DR14 quasars using the spectral fitting code PyQSOFit{\footnote{\url{https://github.com/legolason/PyQSOFit}}} \citep{2018ascl.soft09008G}.
For a detailed description of the emission line fitting, we refer the reader to 
\citetalias{2020ApJS..249...17R}. In short, the SDSS DR14 quasars' spectra obtained from 
the SDSS data server were first corrected for Galactic extinction and brought 
to the rest frame. Spectral modeling was then performed which includes continuum 
and emission line modeling.

The continuum model includes host galaxy subtraction 
based on principal component analysis \citep[PCA;][]{2004AJ....128.2603Y} and 
the AGN continuum model which is a sum of power-law and the Balmer component. The Fe II 
emission in the optical \citep[$3686-7484$~\AA;][]{1992ApJS...80..109B} and 
UV \citep[$1200-3500$~\AA ;][]{2001ApJS..134....1V,2006ApJ...650...57T,2007ApJ...662..131S} were also subtracted during this process. Finally, the AGN emission lines 
were modeled using multiple Gaussians, three Gaussians for broad components 
having FWHM $>900$ km s$^{-1}$, and a single Gaussian for narrow components having 
FWHM $<900$ km s$^{-1}$. \citetalias{2020ApJS..249...17R} catalog provides quality flag on each measurement based on several conditions. A ``quality flag=0'' means the associated measurement is reliable and hence for this work, we used only measurements with ``quality flag=0''. Furthermore, we restricted our analysis to objects with continuum S/N $>5$ pixel$^{-1}$ (i.e., `SN\_RATIO\_CONT$>$5') and peak flux of the broad component of H$\beta$ and Mg II lines larger than 5$\times$ error in peak flux (i.e., `PEAK\_FLUX\_**\_BR$>5\times$`PEAK\_FLUX\_**\_BR\_ERR'). With the above conditions, we arrived at a sample of 36,218 quasars with H$\beta$, 152,425 quasars with Mg II and 25,624 quasars with both H$\beta$ and Mg II line information.

\begin{figure*}
\resizebox{18.2cm}{6cm}{\includegraphics{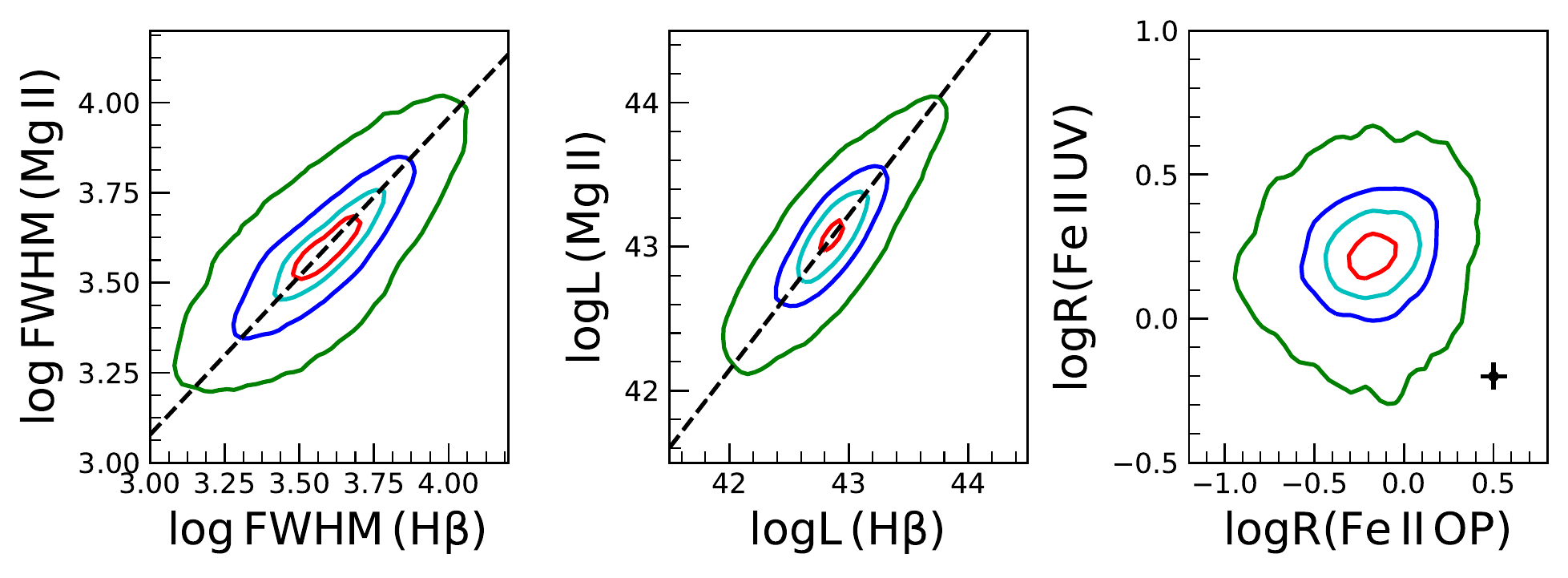}}  
\caption{The relation between the FWHM (left) and luminosity (middle) of the Mg II 
line against the H$\beta$ line. The best fit relation is shown as dashed line. 
The flux ratio of UV Fe II (2200-3090 \AA) to Mg II (R$_{\mathrm{Fe II, UV}}$) 
against the flux ratio of optical Fe II (4435-4685 \AA) to H$\beta$ 
(R$_{\mathrm{Fe II, OP}}$) is shown in the right panel. The black symbol represents average error bar. The 20 (red), 40 (cyan), 68 (blue) and 95 (green) percentile density contours are shown in each panel.}\label{Fig:FWHM_LINE_MGII_HB.png} 
          \end{figure*}

\section{Emission line properties in optical and UV}\label{sec:line_prop}
In Figure \ref{Fig:FWHM_LINE_MGII_HB.png}, we plot several correlations between 
H$\beta$ and Mg II line properties of 25,624 quasars for which both H$\beta$ and Mg II line information is available. We calculated the Spearman rank correlation coefficient ($r_s$) and null-hypothesis probability of no correlation ($p$-value) using Monte Carlo simulations of 10,000 iterations. In each iteration, first, data points were modified by a random Gaussian deviates of zero mean and standard deviation given by the measurement uncertainty and then performed the Spearman rank correlation test \citep[see][]{2014arXiv1411.3816C}. From the distribution of the results, we calculated the median value at 50 percentile and the lower and upper uncertainty at the 16 and 84 percentile, respectively. We find a strong positive correlation between H$\beta$ and Mg II line widths (left panel) with $r_s=0.698^{+0.002}_{-0.002}$ and 
$p<10^{-200}$. All the correlation results presented in this paper are given in Table \ref{Table:coeff}.

 From the linear regression analysis, we found that the line widths 
of Mg II and H$\beta$ lines are related as
\begin{equation}
\mathrm{\log FWHM \,(Mg II)} = \alpha \times \mathrm{\log FWHM \,(H\beta)} + \beta
\end{equation}  
where $\alpha$ and $\beta$ values are given in Table \ref{Table:corr}. For 
this analysis, we used two other methods, (1) LINMIX 
code\footnote{\url{https://github.com/jmeyers314/linmix}} 
\citep{2007ApJ...665.1489K}, which is a Bayesian method using errors on both 
the axes and (2) BECS\footnote{\url{https://github.com/rsnemmen/BCES}} 
\citep{1996ApJ...470..706A,2012Sci...338.1445N} which again is a linear regression 
method using measurement errors on both the axes. The BECS provides four sets of 
measurements which are mentioned in Table \ref{Table:corr}; (a) assuming X as the 
independent variable (Y$|$X), (b) assuming Y as the independent variable (X$|$Y), 
(c) line that bisects the Y$|$X and X$|$Y (Bisector) and (d) line that minimizes 
orthogonal distances (Orthogonal). All the methods provide a similar slope. As it is not clear which variable is independent, we, therefore, adopted the results 
obtained from the BCES Orthogonal method as the best result. However, depending  
on the method used, an upper limit of FWHM (H$\beta$)=2000 km s$^{-1}$ usually used to 
define NLS1 corresponds to a similar FWHM of Mg II line $\sim$2063-2315 km s$^{-1}$ (see Table \ref{Table:corr}). Such a strong correlation between FWHM of H$\beta$ and Mg II line has also been found 
by \citet{2015ApJS..221...35K} with the linear correlation coefficient $r=0.77$. 
Also, \citet{2015ApJ...806..109J} studied line width correlation for high-luminous 
and high-$z$ sources and found that the FWHM of Mg II could be a good substitute 
for FWHM H$\beta$ and this relation was not found to evolve with redshift or 
luminosity. Similarly, we found the luminosities of the Mg II and H$\beta$ lines 
to be strongly correlated (middle panel) having $r_s=0.791^{+0.001}_{-0.001}$ and $p<10^{-200}$ (see Table \ref{Table:coeff}) as expected in a flux-limited sample. We also performed a  linear regression analysis to 
the $\mathrm{\log L \,(Mg II)} - \mathrm{\log L \,(H\beta)}$ diagram 
(Table \ref{Table:corr}). Both LINMIX and BCES (Orthogonal) provided a slope of 
unity.

 \begin{table}
      \caption{Spearman rank correlation analysis. Columns are (1) relation (2) correlation coefficient ($r_s$), (3-4) probability of no correlation ($p$-value) and 1$\sigma$ upper uncertainty  (5) number of data points.}
     	\begin{center}
     	\hspace*{-1cm}
      	\resizebox{1.1\linewidth}{!}{%
          \begin{tabular}{l l l l l}\hline \hline 
          y vs x     &  $r_s$   & $p$ & +$e_p$ & N \\
          (1)    & (2)      & (3)         &  (4)      & (5)    \\\hline
          FWHM (Mg II) $-$  FWHM (H$\beta$)  & $0.698^{+0.002}_{-0.002}$ & $<10^{-200}$   &  $-$      & 25624 \\ 
          L (Mg II) $-$  L (H$\beta$) & $0.791^{+0.001}_{-0.001}$ & $<10^{-200}$   & $-$ & 25624 \\
          R(Fe II, UV) $-$ R(Fe II, OP) & $0.114^{+0.005}_{-0.005}$ & ${2\times10^{-29}}$ & $5\times10^{-27}$ & 9584 \\
          FWHM (H$\beta$) $-$ EW (H$\beta$) & $0.130^{+0.004}_{-0.004}$ & $6\times10^{-98}$ & $7\times 10^{-92}$ & 25624 \\ 
          FWHM(H$\beta$) $-$ EW (Fe II, OP) & $-0.294^{+0.005}_{-0.005}$ & $1\times 10^{-190}$ & $2\times10^{-184}$ & 9584 \\
          FWHM (Mg II) $-$ EW (Mg II)       & $0.347^{+0.003}_{-0.003}$ & $<10^{-200}$ & $-$ & 25624 \\
          FWHM (Mg II) $-$ EW (Fe II, UV)   & $0.143^{+0.004}_{-0.004}$ & $5\times 10 ^{-45}$  &  $9\times10^{-43}$   & 9584\\
          
           \hline \hline
           			  
             \end{tabular} } 
             \label{Table:coeff}
             \end{center}
         \end{table}

 \begin{table}
      \caption{Results of the correlation analysis (y = $\alpha$x+ $\beta$)
 between Mg II and H$\beta$ lines. The columns are as follows (1) y-axis variable, (2) x-axis variable, (3) method used to perform linear fit and (4)-(5) slope and intercept of the linear fit.}
     	\begin{center}
     	\hspace*{-1cm}
      	\resizebox{1.1\linewidth}{!}{%
          \begin{tabular}{ l l l l l}\hline \hline 
          y   &  x   & method & $\alpha$ & $\beta$ \\
          (1) & (2)  & (3)    & (4)      & (5)   \\\hline
          log FWHM(Mg II) & log FWHM (H$\beta$) & LINMIX         & $0.792 \pm 0.004$ &  $0.750\pm 0.012$ \\
           	           & 				     & BCES (Y$|$X)      & $0.812 \pm 0.005$ &  $0.681\pm 0.018$ \\
           			   &				     & BCES (Y$|$X)		 & $0.975 \pm 0.007$ &  $0.096\pm 0.026$ \\
           			   &				     & BCES (Bisector)   & $0.890 \pm 0.005$ &  $0.401\pm 0.017$ \\
           			   &				     & BCES (Orthogonal) & $0.880 \pm 0.005$ &  $0.438\pm 0.018$ \\ 
           			    
           log L(Mg II)   & log L (H$\beta$)    & LINMIX         & $0.888 \pm 0.004$ &  $4.998\pm 0.159$ \\ 
           			   &					 & BCES (Y$|$X)      & $0.897 \pm 0.004$ &  $4.593\pm 0.171$ \\ 
           			   &					 & BCES (X$|$Y)      & $1.258 \pm 0.006$ &  $-10.881\pm 0.270$ \\
           			   &					 & BCES (Bisector)   & $1.062 \pm 0.004$ &  $-2.452\pm 0.174$ \\ 
           			   &					 & BCES (Orthogonal) & $1.074 \pm 0.005$ &  $-3.006\pm 0.213$ \\  
           \hline \hline
           			  
             \end{tabular} } 
             \label{Table:corr}
             \end{center}
         \end{table}

NLS1 shows stronger Fe II emission in the optical compared to the BLS1 \citep{2017ApJS..229...39R}. To find out any correlation between optical and UV Fe II strength, we plot the optical Fe II strength ($R_{\mathrm{FeII, OP}}$) which is the 
ratio of the EW of Fe II in the wavelength range of $4435-4685$~\AA 
\, to the H$\beta$ against the UV Fe II strength ($R_{\mathrm{FeII, UV}}$) which 
is the ratio of the EW of Fe II in the wavelength range $2200-3090$~\AA\, 
to the Mg II in the right panel of Figure \ref{Fig:FWHM_LINE_MGII_HB.png}. A weak 
but positive correlation between them with $r_s=0.114^{+0.005}_{-0.005}$ and $p=2\times 10^{-29}$ is present albeit with large dispersion.

\begin{figure}
          \resizebox{9cm}{8cm}{\includegraphics{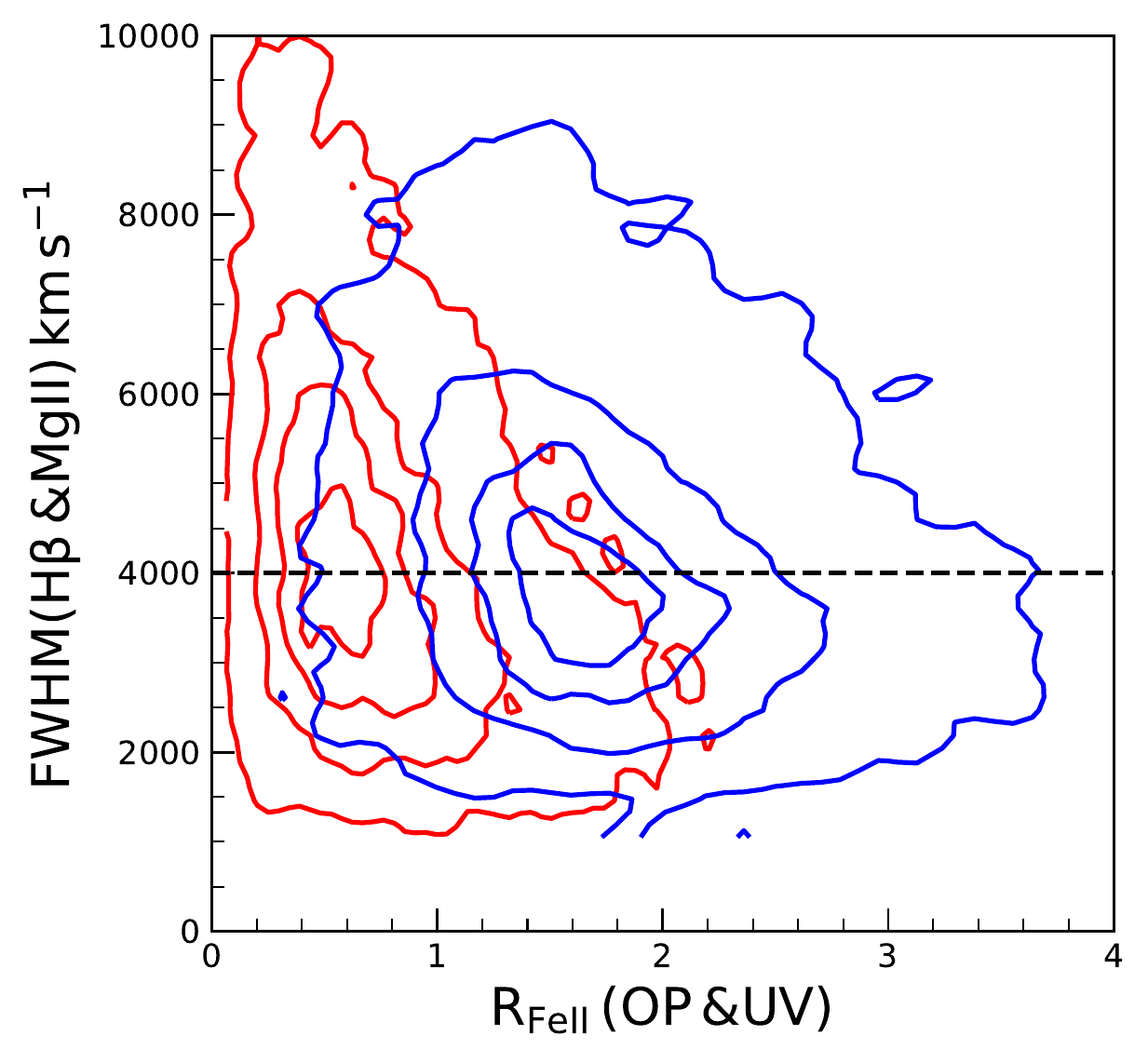}}  
          \caption{Relation between the line width of H$\beta$ against the 
optical Fe II strength R$_{\mathrm{Fe II, OP}}$ (red contours) and Mg II line 
width against the UV Fe II strength R$_{\mathrm{Fe II, UV}}$ (blue). The 20, 40, 68 and 95 percentile density contours are shown from inner to outer.}\label{Fig:RFEII_FWHM_OP_UV} 
          \end{figure}

From principal component analysis (PCA) of a sample of PG quasars \citet{1992ApJS...80..109B} found that the first eigenvector (EV1), which explains the difference between various AGN types, is dominated by the strong anti-correlation between the strength of Fe II and [O III]5007 as well as radio-loudness, H$\beta$ FWHM and asymmetry. The optical plane of EV1 correlation involves two main parameters, FWHM of H$\beta$ and the strength of Fe II to H$\beta$ \citep{2000ARA&A..38..521S}. The main driver of EV1 is believed to be the orientation and Eddington ratio \citep{2014Natur.513..210S}. The 
former strongly affects the line kinematics and particularly important for objects 
having flat broad line region (BLR) geometry while the latter is found to be strongly correlated with 
the Fe II strength. The NLS1s, by definition, are located at the lower part of the EV1 diagram \citep[e.g.,][]{2017ApJS..229...39R}. To find out any similarities of the optical plane of EV1 with the UV plane of EV1, in Figure \ref{Fig:RFEII_FWHM_OP_UV}, we plot line FWHM 
against $R_{\mathrm{FeII}}$ in the optical and UV. The FWHM (H$\beta$) v/s. 
$R_{\mathrm{FeII, OP}}$ is shown in red while FWHM (Mg II) v/s $R_{\mathrm{FeII, UV}}$ is shown in blue. We do see a similar shape in both optical and UV EV1 diagrams.

\begin{figure}
          \resizebox{9cm}{9cm}{\includegraphics{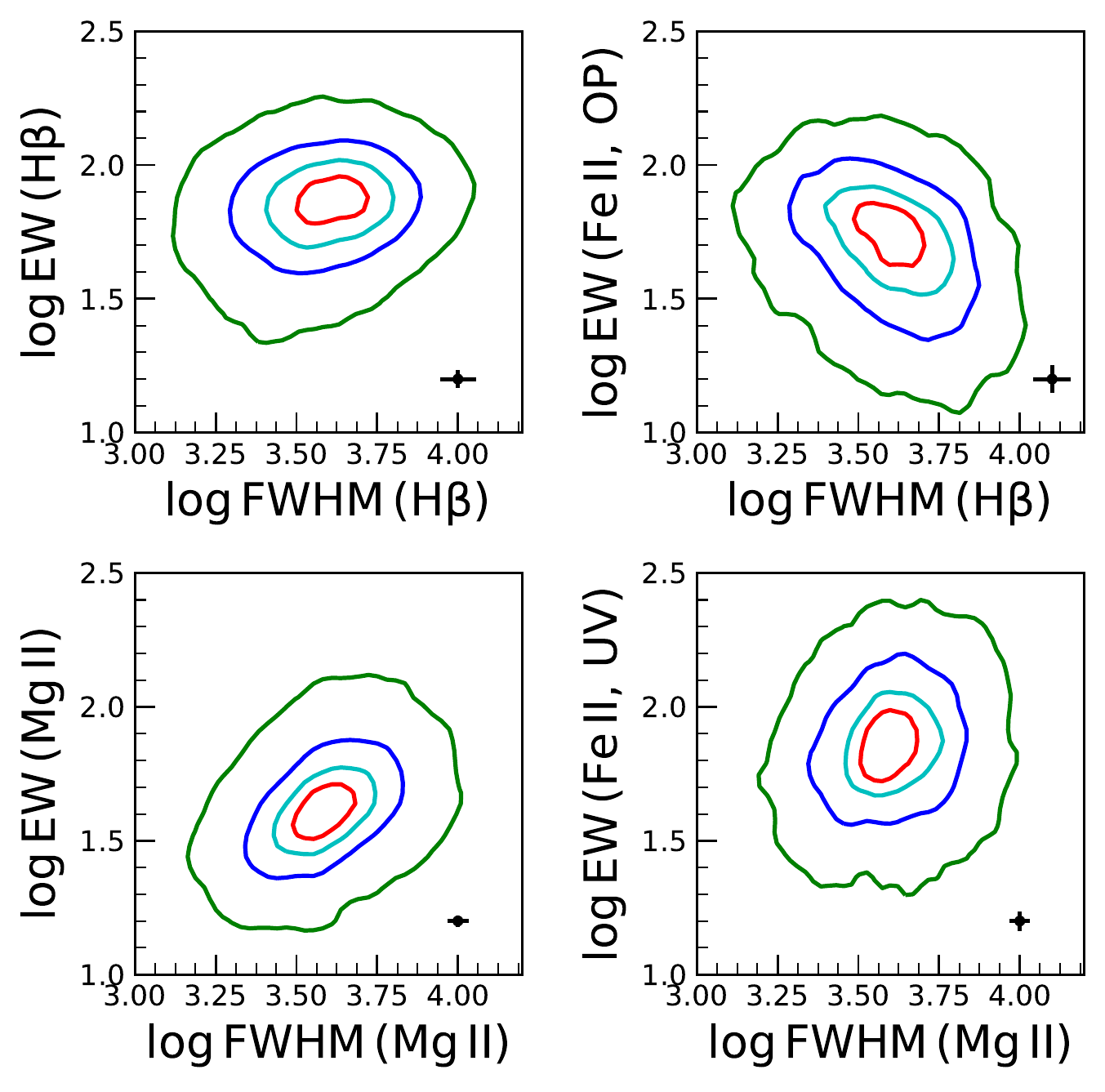}}  
          \caption{Top panels: H$\beta$ and Optical Fe II line EW are plotted against the FWHM of H$\beta$. Bottom panels: Mg II and UV Fe II line EW are plotted against the FWHM of Mg II. The black symbol in each panel represents the average error bar. The 20 (red), 40 (cyan), 68 (blue) and 95 (green) percentile density contours are shown.}\label{Fig:EW} 
          \end{figure}

As shown in Figure \ref{Fig:EW}, the shape of EV1 in UV is dominated by the 
strong positive correlation ($r_s=0.347^{+0.003}_{-0.003}$) between the FWHM and  
 EW of Mg II, since the EW of UV Fe II is weakly correlated ($r_s=0.143^{+0.004}_{-0.004}$) with the FWHM of Mg II (see Table \ref{Table:coeff}). While in the optical, the FWHM of H$\beta$ is weekly correlated ($r_s=0.130^{+0.004}_{-0.004}$) with its EW,  the EW of Fe II in the optical is anti-correlated ($r_s=-0.294^{+0.005}_{-0.005}$) with the FWHM of H$\beta$. Such differences between UV 
and optical Fe II lines have been noticed by \citet{2015ApJS..221...35K}. The 
differences could be due to different spatial distribution of clouds in the 
emitting region and/or the different excitation mechanisms. For example, 
\citet{Ferland_2009} suggested that the cloud distribution of the UV Fe II is asymmetric 
while for optical Fe II line they are isotropic \citep[see][]{2011MNRAS.410.1018S}. 
Indeed \citet{2015ApJS..221...35K} found that although the UV and optical lines 
originate around the similar region in AGN,  The UV Fe II emitting clouds have asymmetric distribution while the optical Fe II clouds have isotropic 
distribution. On the other hand, different excitation mechanisms could be 
responsible for the optical and UV Fe II emission. The optical depth of UV Fe II 
is larger than the optical Fe II, hence  UV Fe II lines decrease more rapidly 
than the optical Fe II with the increase of column 
density \citep{1987A&A...184...33J}.

\section{high-z NLS1 candidates}\label{sec:high-z-nls1}
Since NLS1s are classified based on the criteria of FWHM(H$\beta) <2000$ km s$^{-1}$, 
and [O III]/H$\beta <3$, one can use the correlation between Mg II and H$\beta$ 
line width to select high-$z$ NLS1s when the H$\beta$ line is not present. We, therefore, 
selected an object to be a NLS1 if the FWHM of the Mg II line is $<2000$ km s$^{-1}$\footnote{FWHM(H$\beta$)=2000 km s$^{-1}$ corresponds to FWHM(Mg II) of $\sim$2063-2315 km s$^{-1}$ depending on the choice of correlation (see Table \ref{Table:corr}), however, to be conservative, we adopted FWHM(Mg II) $<2000$ km s$^{-1}$ as the criterion to define high-z NLS1.}. We note that the classical definition of NLS1 is subjective and has been debated since the broad line widths do not show any bimodality or discontinuity. A more conservative limit at 
FWHM(H$\beta$)=4000 km s$^{-1}$ has been suggested where sources below 4000 km s$^{-1}$ are termed as 
population A while above 4000 km s$^{-1}$ are termed as population B based on the 
optical plane of EV1 diagram \citep{2000ARA&A..38..521S,2018FrASS...5....6M}. Here, 
NLS1s fall under population A. 
Moreover, many NLS1s are found to be very weak Fe II emitters. Thus, 
\citet{2006ApJS..166..128Z} and \citet{2017ApJS..229...39R} used a cutoff of 
2200 km s$^{-1}$ slightly larger than 2000 km s$^{-1}$. A larger FWHM increases the 
sample, however, the average black hole mass of the same also increases. On the 
other hand \citet{2007ApJ...654..754N} suggested using the Eddington ratio 
cutoff of $L/L_{\mathrm{Edd}} \ge0.25$.

The [O III]/H$\beta <3$ criterion mainly separate NLS1s from Type 2 AGN\footnote{IRAS 20181-2244, has a flux ratio of [O III] to H$\beta >3$ but the object is classified as NLS1 type by \citet{1998ApJ...494..194H}.} in the 
low-resolution and low-SNR spectra since line fluxes are difficult to measure \citep[see][]{1999MNRAS.307L..47H,1998ApJ...501..103H}.  
This is not the case in our study as we restricted our sample with strong emission lines. 
Also, this flux ratio criterion has been 
excluded when high ionization iron lines, e.g., [Fe vii] k6087 and [Fe x] k6375 
are present \citep[exception, for example, IC 3599;][]{2001A&A...372..730V}. Since 
all of our objects have broad line width larger than 900 km s$^{-1}$, larger than 
Type 2 AGN and are quasars based on the absolute magnitude, we ignore the flux 
ratio criterion. With the above conditions, we obtain a sample of 2684 high-$z$ 
NLS1 candidates\footnote{About 11\% of NLS1 candidates in our sample have a fractional error in the FWHM of Mg II broad component larger than unity i.e., FWHM\_MGII\_BR\_ERR$>$FWHM\_MGII\_BR, these objects should be considered with caution.}. The catalog of high-$z$ NLS1 candidates is available on Zenodo (\href{https://doi.org/10.5281/zenodo.4405039}{doi:10.5281/zenodo.4405039}) and column information are given in Table \ref{Table:catalog}. Note that measurements are directly taken from \citetalias{2020ApJS..249...17R}.

To study the high-$z$ NLS1 properties, we also compare it with the low-$z$ 
NLS1 sample ($z<0.8$) present in the DR 14 quasar catalog. This low-$z$ sample, 
which is selected as per the classical definition of FWHM (H$\beta$) $<2000$ km s$^{-1}$ 
and [O III]/H$\beta <3$ has 3109 NLS1. If we exclude the flux ratio criterion, we 
found 3131 sources. This suggests the 
flux ratio criterion is not important in our selection of high-$z$ NLS1s.

 \begin{figure}
 \resizebox{9cm}{7cm}{\includegraphics{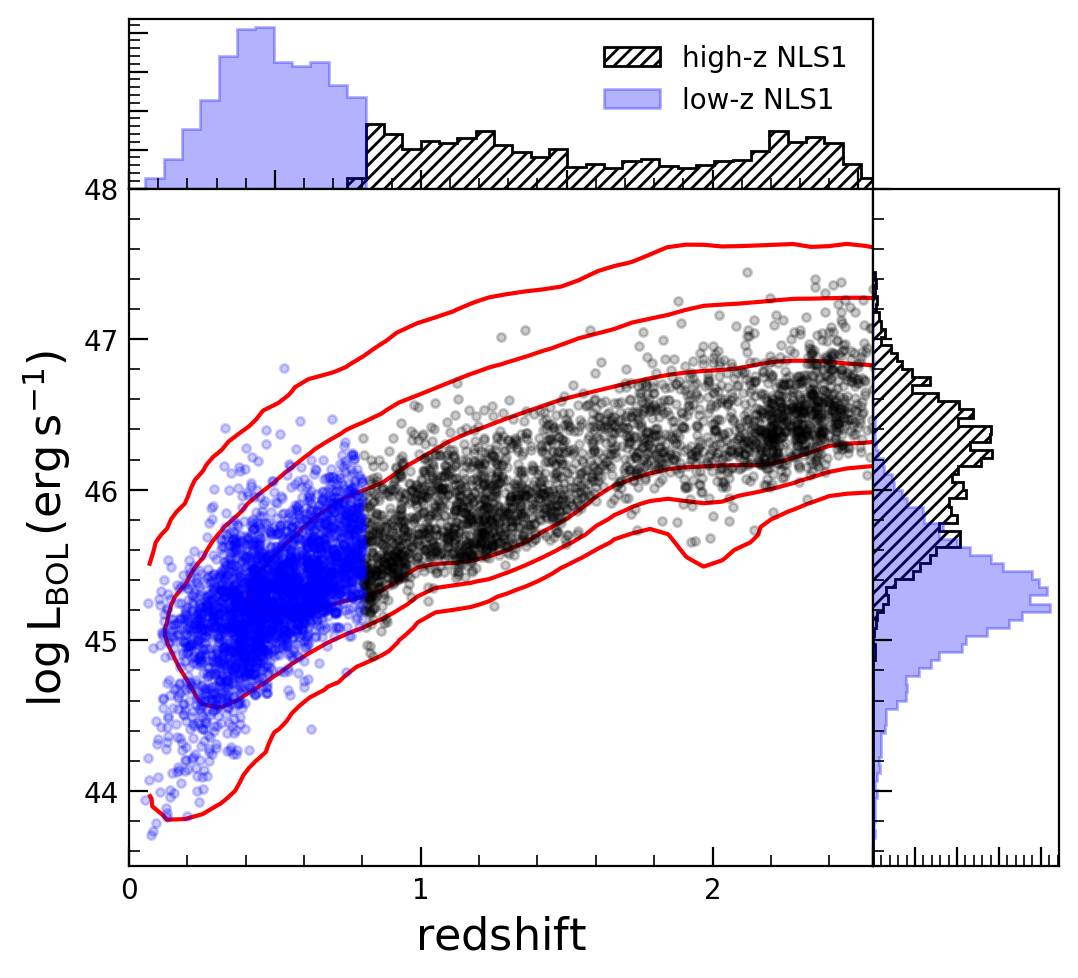}}
 \caption{The bolometric luminosity against redshift. The low-z NLS1 (blue) and high-z NLS1 candidates (black) are shown along with the parent sample (1$\sigma$, 2$\sigma$, 3$\sigma$ density contours) of SDSS DR14 quasars. The top and right panels show the distribution of redshift and bolometric luminosity, respectively for NLS1.}\label{Fig:L3000} 
 \end{figure}

       \begin{figure}
      \resizebox{9cm}{8cm}{\includegraphics{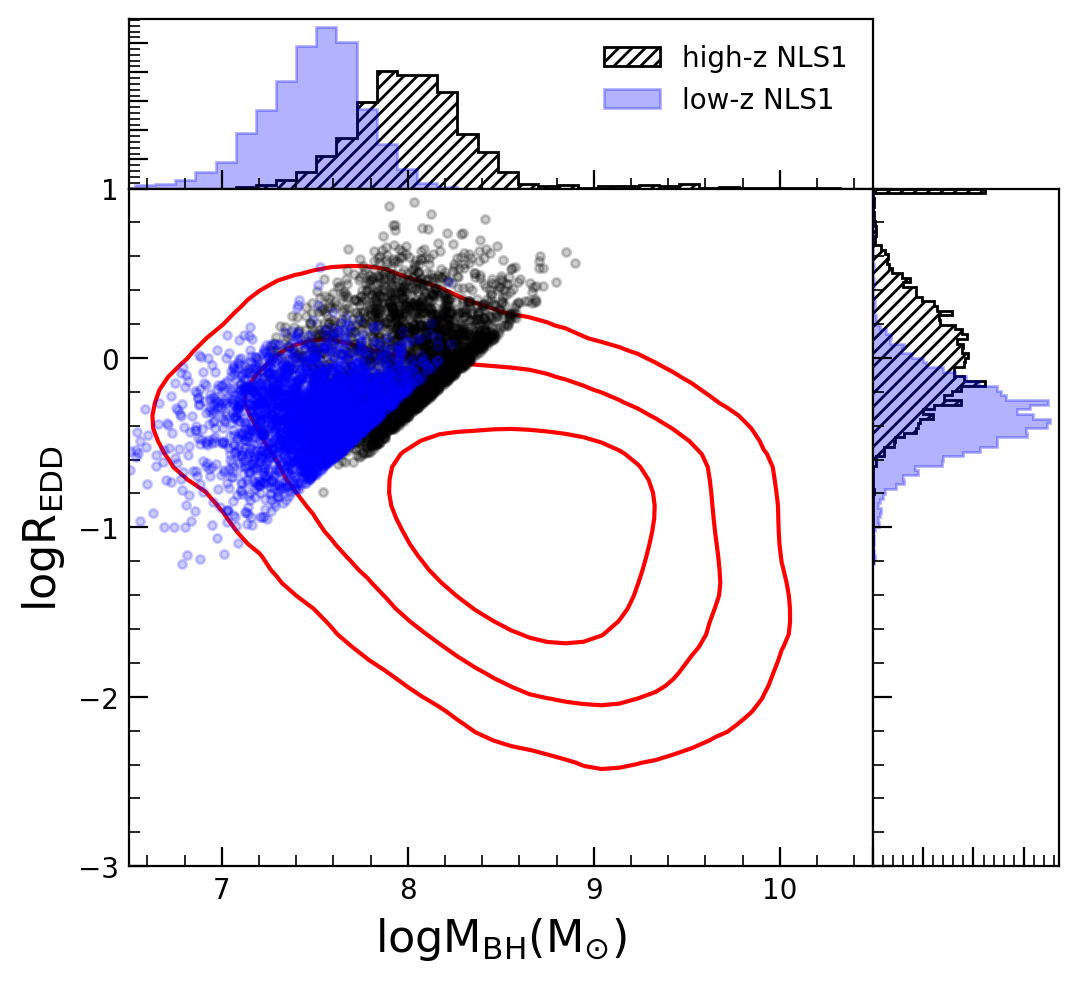}}  
      \caption{The Eddington ratio is plotted against black hole mass for 
low-z NLS1 (blue), high-z NLS1 candidates (black dots) along with the SDSS DR14 
quasars for $z<2.2$ (1$\sigma$, 2$\sigma$, and 3$\sigma$ density contours). The histograms of black hole mass (top panel) and Eddington ratio 
(right panel) are also shown.}\label{Fig:MBH_REDD} 
      \end{figure}

  \begin{table*}
       \caption{High-z NLS1 catalog. Columns are (1) FITS column number, (2) name of the column, (3) format (4) unit, and (5) description. All the quantities are directly taken from \citetalias{2020ApJS..249...17R}.}
      	\begin{center}
      	\hspace*{-1cm}
       	\resizebox{1.1\linewidth}{!}{%
           \begin{tabular}{lllll}\hline \hline  
    Number & Column Name              & Format    & Unit  & Description \\ 
    (1)    &  (2)                     & (3)       & (4)   &   (5)       \\ \hline
 1	&	SDSS\_NAME					&	String	&	     		& \pbox{40cm}{Object name as given in \citetalias{2020ApJS..249...17R}.} 		\\
 2	&	RA							&	Double	& Degree 		& Right Ascension (J2000)		\\
 3	&	DEC							&	Double	& Degree 		& Declination (J2000)	\\
 4	&	SDSS\_ID					&	String	&				& PLATE-MJD-FIBER	\\
 5	&	REDSHIFT					&	Double	&				& Redshift	\\
 6	&   LOG\_L3000					&	Double	& erg s$^{-1}$	& Logarithmic continuum luminosity at rest-frame 3000 \AA	\\
 7	&   LOG\_L3000\_ERR				&	Double	& erg s$^{-1}$	& Measurement error in LOG\_L3000	\\
 8	&   FWHM\_MGII\_BR				&	Double	& km s$^{-1}$	& FWHM of Mg II broad component 	\\
 9	&   FWHM\_MGII\_BR\_ERR			&	Double	& km s$^{-1}$	& Measurement error in FWHM\_MGII\_BR	\\
 10	&   EW\_MGII\_BR				&	Double	& \AA 			& Rest-frame equivalent width of Mg II broad component	\\
 11	&   EW\_MGII\_BR\_ERR			&	Double	& \AA 			& Measurement error in EW\_MGII\_BR	\\
 12	&   LOGL\_MGII\_BR				&	Double	& erg s$^{-1}$	& Logarithmic line luminosity of Mg II broad component	\\
 13	&   LOGL\_MGII\_BR\_ERR			&	Double	& erg s$^{-1}$	& Measurement error in LOGL\_MGII\_BR	\\
 14	&   LOGL\_MGII\_NA				&	Double	& erg s$^{-1}$	& Logarithmic line luminosity of Mg II narrow component	\\
 15	&   LOGL\_MGII\_NA\_ERR			&	Double	& erg s$^{-1}$	& Measurement error in LOGL\_MGII\_NA	\\
 16	&   LOGL\_FE\_UV				&	Double	& erg s$^{-1}$	& \pbox{40cm}{Logarithmic luminosity of the UV Fe II complex \\ within the 2200-3090 \AA}	\\
 17	&   LOGL\_FE\_UV\_ERR			&	Double	& erg s$^{-1}$	& Measurement error in LOGL\_FE\_UV	\\
 18	&   EW\_FE\_UV					&	Double	& \AA			& \pbox{40cm}{Rest-frame equivalent width of UV Fe II complex \\ within the 2200-3090 \AA}	\\
 19	&   EW\_FE\_UV\_ERR				&	Double	& \AA			& Measurement error in EW\_FE\_UV	\\
 20	&   LOG\_MBH					&	Double	& $M_{\odot}$	& Logarithmic fiducial single-epoch BH mass	\\
 21	&   LOG\_MBH\_ERR				&	Double	& $M_{\odot}$	& Measurement error in LOG\_MBH	\\
 22	&   LOG\_REDD					&	Double	&				& \pbox{40cm}{Logarithmic Eddington  ratio based on \\fiducial single-epoch BH mass}	\\
  \hline \hline
     \end{tabular} } 
    \label{Table:catalog}
     \end{center}
   \end{table*}

 \begin{table}
      \caption{Spearman rank correlation analysis. Columns are (1) relation (2) sample (3) name of data points (4) correlation coefficient ($r_s$), (5-6) probability of no correlation ($p$-value) and 1$\sigma$ upper uncertainty.}
     	\begin{center}
     	\hspace*{-1cm}
      	\resizebox{1.1\linewidth}{!}{%
          \begin{tabular}{l l l l l l}\hline \hline 
          y vs x  & sample  &  N          &  $r_s$    & $p$          & +$e_p$ \\
          (1)    & (2)      & (3)         &  (4)      & (5)          & (6)   \\\hline
          R(Fe II, UV) vs. $R_{\mathrm{EDD}}$ & high-z NLS1 & 2002   & $0.077^{+0.019}_{-0.019}$ & $5 \times 10^{-4}$ & $9 \times 10^{-3}$ \\
          R(Fe II, OP) vs. $R_{\mathrm{EDD}}$ & low-z NLS1  & 2917   & $0.095^{+0.014}_{-0.014}$ & $2\times10^{-7}$ & $9\times10^{-6}$ \\
          R(Fe II, UV) vs. $R_{\mathrm{EDD}}$ & DR14 quasars & 115354 & $0.237^{+0.001}_{-0.001}$ & $<10^{-200}$ & -- \\
          R(Fe II, OP) vs. $R_{\mathrm{EDD}}$ & DR14 quasars & 28577 & $0.361^{+0.002}_{-0.002}$ & $<10^{-200}$ & -- \\
          EW(Fe II, OP)/EW(Fe II, UV) vs. $R_{\mathrm{EDD}}$ & DR14 quasars & 9584 & $0.282^{+0.004}_{-0.004}$ & $3\times10^{-175}$ & $9\times10^{-170}$ \\
          $R_{1.4}$ vs. $L_{1.4}$  &  high-z NLS1 & 130  & $0.800^{+0.003}_{-0.003}$ & $3\times10^{-30}$ & $3\times5^{-30}$ \\
          $R_{1.4}$ vs. $L_{1.4}$  &  low-z NLS1 & 187   &  $0.727^{+0.005}_{-0.005}$ & $4\times10^{-32}$ & $1\times 10^{-31}$ \\
          $R_{1.4}$ vs. $L_{1.4}$  &  DR14 quasars & 18236  & $0.778^{+0.001}_{-0.001}$ & $<10^{-200}$ & $--$ \\
          $M_{BH}$ vs. $L_{1.4}$   & high-z NLS1 & 130 & $0.394^{+0.048}_{-0.049}$ & $3\times10^{-6}$ & $5\times10^{-5}$ \\  
          $M_{BH}$ vs. $L_{1.4}$   & low-z NLS1 & 187 & $0.313^{+0.042}_{-0.043}$ & $1\times10^{-5}$ & $2\times10^{-4}$ \\
          $M_{BH}$ vs. $L_{1.4}$   & All NLS1 & 317 & $0.596^{+0.023}_{-0.022}$ & $5\times10^{-32}$ & $3\times10^{-29}$ \\
          $M_{BH}$ vs. $L_{1.4}$   & DR14 quasars ($z<2.2$) & 10437 & $0.218^{+0.003}_{-0.003}$ & $5\times10^{-113}$ & $1\times10^{-109}$ \\

           \hline \hline
             \end{tabular} } 
             \label{Table:coeff2}
             \end{center}
         \end{table}

\section{Properties of high-z NLS1}\label{sec:prop}
We used the values of the bolometric luminosities from \citetalias{2020ApJS..249...17R}, 
who measured them from the monochromatic luminosities after applying the 
corresponding bolometric corrections. In Figure \ref{Fig:L3000}, we plot 
the bolometric luminosities 
against redshifts for the low-$z$ NLS1s (blue) and high-$z$ NLS1 candidates (black). 
In the same figure, the density contours of SDSS DR14 quasars are also shown. The 
median $\log L_{\mathrm{bol}}$ for low-$z$ and high-$z$ NLS1 candidates are found 
to be $45.29\pm0.39$ erg s$^{-1}$\footnote{Here uncertainties are 1$\sigma$ dispersion of the associated quantity.} and $46.16\pm0.42$ erg s$^{-1}$, respectively. 
Their redshift distribution ranges from $0.06-0.8$ for low-z NLS1s and $0.8-2.57$ 
for high-$z$ NLS1s.

\citetalias{2020ApJS..249...17R} provided fiducial virial black hole masses based on 
H$\beta$ (for $z < 0.8$), Mg II (for $0.8 \le z < 1.9$) and CIV (for $z \ge 1.9$) 
lines depending on the redshift with decreasing preference from H$\beta$ to CIV. 
They also estimated the Eddington ratio, which is the ratio of bolometric to 
Eddington luminosity where the former is based on the continuum luminosity with the 
appropriate bolometric correction factor and the latter is based on the black hole 
masses. We used the measurements provided in \citetalias{2020ApJS..249...17R}. Note that 
for $z>1.9$ sources, the fiducial virial masses are based on  CIV line and  
they tend to have larger uncertainties than Mg II masses \citep[e.g.,][]{2012ApJ...759...44D,2012MNRAS.427.3081T,2017ApJ...839...93P,2020ApJS..249...17R}. Hence many of the high-$z$ NLS1 in the redshift range of $z>1.9$, for which 
\citetalias{2020ApJS..249...17R} provides Mg II based masses, we used them instead of 
the fiducial masses.

In Figure \ref{Fig:MBH_REDD}, we plot the distribution of $M_{\mathrm{BH}}$ values and
Eddington ratios for the low-z NLS1s (blue-dots) and high-$z$ 
NLS1 candidates (black-dots). They are also shown in the $M_{\mathrm{BH}}$ versus
Eddington ratio plane. We also plot the density contours (1$\sigma$, 2$\sigma$, 
3$\sigma$) of SDSS DR14 quasars for $z<2.2$. The NLS1s are 
located at the extreme upper left corner of the diagram having low black hole mass 
and higher Eddington ratio. The median logarithmic black hole mass of high-$z$ NLS1 
candidates is $8.01\pm0.35 \, M_{\odot}$  slightly larger than the sample of 
low-z NLS1s ($7.51\pm0.27\, M_{\odot}$). The higher mass of the high-z NLS1 sample is due to the higher luminosity compared to the low-z NLS1 sample. However, this is much lower compared to the 
overall sample of SDSS DR14 quasars up to $z<2.2$, which is $8.64\pm0.50\, M_{\odot}$.
Similarly, the logarithmic Eddington ratio distribution has a median of 
$0.02\pm0.27$ and $-0.35\pm0.21$ for high-$z$ and low-$z$ 
NLS1s, respectively. These values are much higher than that of SDSS DR14 quasars 
that have a median value of $-0.96\pm0.43$ up to $z<2.2$.

\begin{figure}
  \resizebox{9cm}{6cm}{\includegraphics{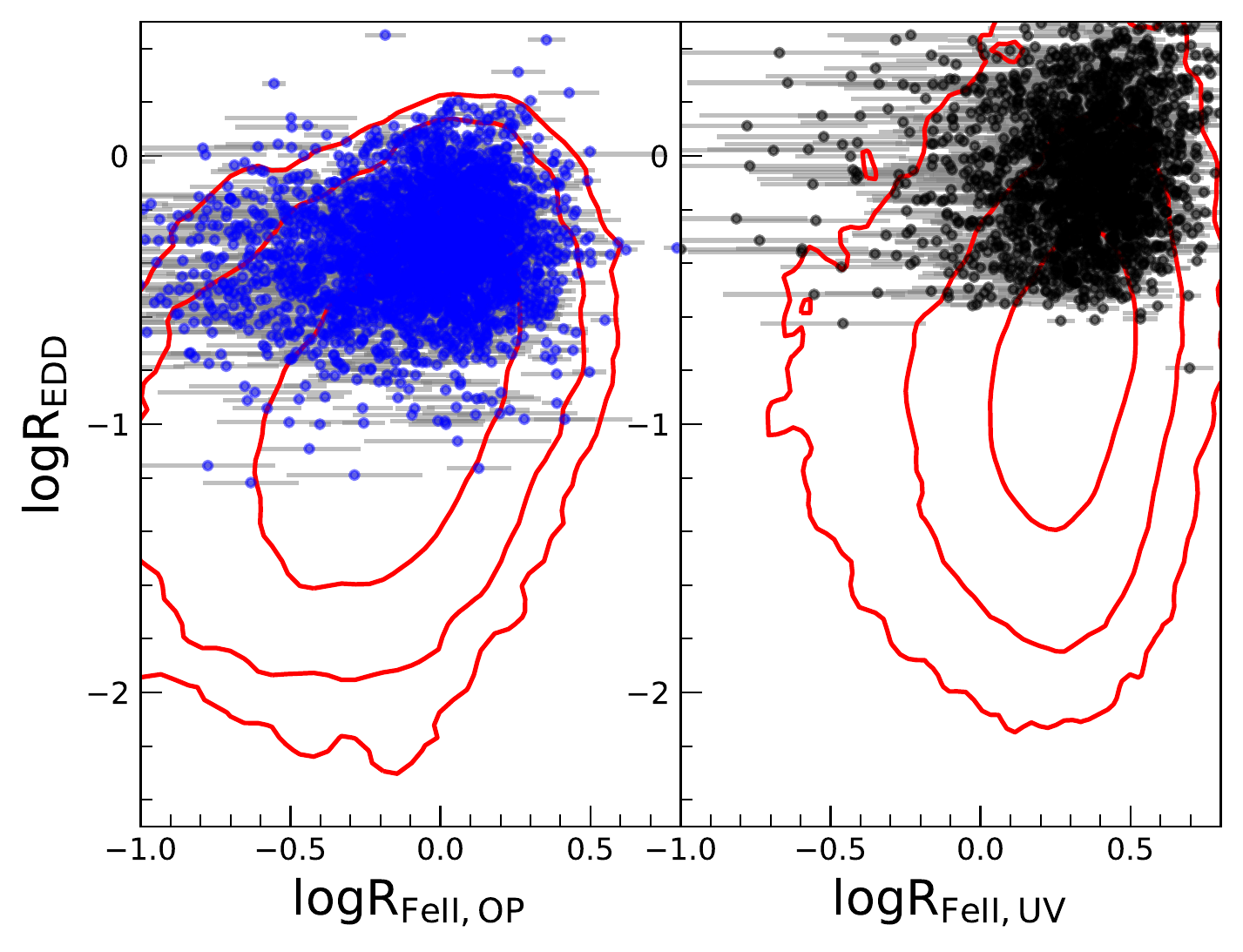}}  
  \caption{Eddington ratio is plotted against R$_{\mathrm{Fe II, OP}}$ (left) while R$_{\mathrm{Fe II, UV}}$ (blue). A positive correlation is noticed. The SDSS DR14 quasars are also shown by the solid contours (1$\sigma$, 2$\sigma$, and 3$\sigma$). }\label{Fig:RFEII_EDD} 
    \end{figure}

In Figure \ref{Fig:RFEII_EDD}, we plot the Eddington ratio as a function of 
Fe II strength in optical (left) and UV (right). We also plotted the density contours of 
SDSS DR14 quasars. The NLS1 candidates are located 
at the extreme upper right corner of the diagram. A moderately strong correlation 
between Fe II strength and Eddington ratio is found having $r_s=0.361^{+0.002}_{-0.002}$ in the optical 
and $0.237^{+0.001}_{-0.001}$ in the UV for SDSS DR14 quasars (see Table \ref{Table:coeff2}). This positive correlation is also present in the low-z and high-z NLS1 samples but very weak due to the small range of the Eddington ratio of these samples. A positive correlation between Fe II strength and Eddington ratio is known to exist  e.g., 
\citet{2018A&A...620A.118N} using a sample of 302 high accreting sources found a 
positive correlation with a correlation coefficient of $0.49$ between the optical Fe II strength and 
Eddington ratio 
similar to what has been found here. Moreover, using a SDSS sample, 
\citet{2019ApJ...874...22S} found a positive correlation with a correlation coefficient of 0.48 between UV Fe II 
strength and Eddington ratio. In fact, the ratio of the equivalent width of optical to UV Fe II increases 
with the Eddington ratio having $r_s=0.282^{+0.004}_{-0.004}$ considering the SDSS DR14 quasars. Such a 
positive correlation but somewhat stronger has been found by 
\citet{2011MNRAS.410.1018S} using a sample of about 900 SDSS quasars from SDSS DR5. 
A dense medium and large column density is found for high accreting sources since 
lower density and small column density clouds could be blown away due to strong 
radiation pressure \citep[see][]{2009ApJ...703L...1D}. However, 
\citet{2004ApJ...614..558N} suggested that increasing star formation could also 
lead to high density and large column density region. A high star formation rate 
in high Eddington ratio and low mass sources which are typically NLS1s is indeed 
found by \citet{2010MNRAS.403.1246S}.

           \begin{figure}
          \resizebox{9cm}{8cm}{\includegraphics{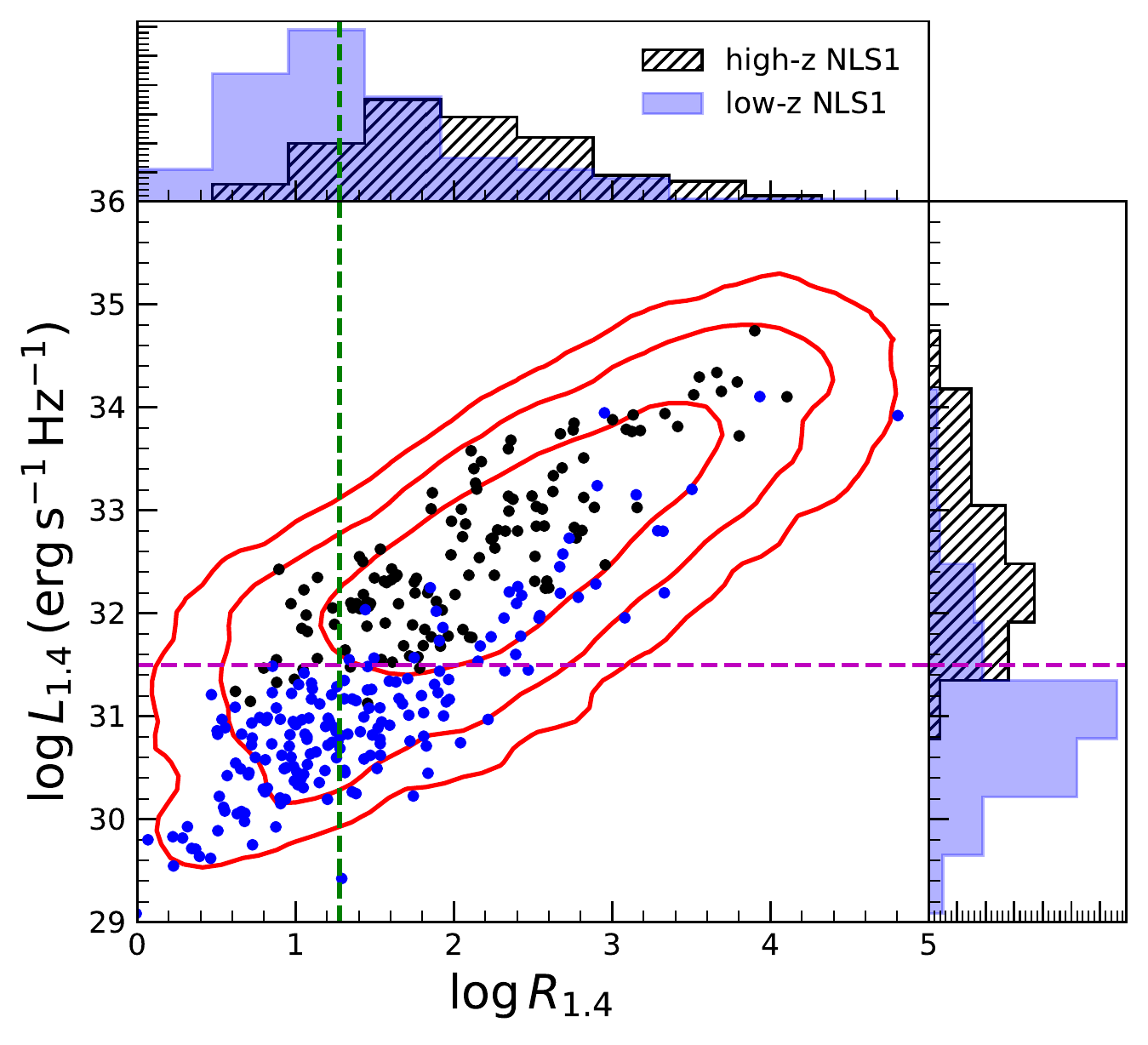}}  
          \caption{The 1.4 GHz radio luminosity against radio loudness. 
The blue and black circles are the low-$z$ and high-$z$
NLS1 candidates respectively.
The 1$\sigma$, 2$\sigma$, and 3$\sigma$ density contours of DR14 quasars are 
shown in red. The vertical line at $R_{1.4}=19$ is the dividing line for the 
radio-quiet and radio-loud quasars while the horizontal line is the separation 
for FR I and II objects. The distributions of the radio loudness parameter
and the radio luminosity at 1.4 GHz are also shown.}\label{Fig:LUM_RADIO} 
        \end{figure}

\subsection{Radio properties}
NLS1s are usually radio-quiet having a low radio detection fraction of 
about 5\% \citep{2006AJ....132..531K,2017ApJS..229...39R,2018MNRAS.480.1796S}. The 
radio-loud fraction is even smaller, about 2.5\%. A fraction of these sources are 
found to have large kpc-scale jet as well as high energy $\gamma$-ray emission 
\citep[see][and references therein]{2018ApJ...853L...2P}. The radio-detection 
fraction of NLS1s is much lower compared to the fraction of typical quasars \citep{1989AJ.....98.1195K}. \citet{2018A&A...613A..51P} 
cross-matched SDSS DR14 quasars with the 1.4GHz FIRST radio survey \citep[catalog 
December 2014;][]{1995ApJ...450..559B} with a radius of 2$^{\prime\prime}$ finding 
a total of 18,273 counterparts. This radio detection fraction, only 4\% of SDSS 
DR14 quasars, is much smaller than the 16\% found in earlier studies 
\citep[e.g.,][]{1989AJ.....98.1195K}. If we consider sources with continuum  
S/N $>5$ pixel$^{-1}$, a similar low radio detection is found, only 6\%. However, 
this fraction is limited by the sensitivity of the FIRST survey which is 1 mJy/beam. 
Indeed, at low redshift $z<0.3$, the detection fraction is found to be 16\% as 
usually considered, however, it decreases to 8\% for $z<0.8$. The detection 
fraction of our low-$z$ NLS1 sample is 7\% while for the high-z NLS1 candidate sample it  
is 6\%. Thus, the fraction of radio detected NLS1s is similar to the SDSS DR 14 quasars fraction and strongly depends on the redshift and the flux limit of the radio survey.

To study the radio properties of the high-redshift NLS1 candidates, we calculated 
the 1.4 GHz luminosity ($L_{\mathrm{1.4GHz}}$) from the integrated sources flux after k-correction assuming a radio spectral index $\alpha=0$ i.e the measured radio flux divided by (1+z) \citep[see][]{2009MNRAS.392..617D}. The radio loudness $R_{\mathrm{1.4GHz}}$ was calculated by the ratio of integrated radio flux to the optical SDSS g-band flux. In Figure \ref{Fig:LUM_RADIO} we plot $L_{\mathrm{1.4GHz}}$ against $R_{\mathrm{1.4GHz}}$. 
The distribution of low-$z$ and high-$z$ NLS1s along both the axes are shown. 
The vertical line represents the division of FR I and FR II sources 
\citep{1974MNRAS.167P..31F}, while the horizontal line at $R_{\mathrm{1.4GHz}}=19$ 
represents the dividing line between radio-quiet and radio-loud sources based on 
the 1.4GHz radio flux \citep{2006AJ....132..531K,2018MNRAS.480.1796S} which 
corresponds to the original dividing line of $R=10$ based on the 5 GHz radio 
flux \citep{1989AJ.....98.1195K}.

Figure \ref{Fig:LUM_RADIO} shows a strong positive correlation ($r_s=0.778^{+0.001}_{-0.001}$) between radio luminosity 
and the radio loudness which is expected as the latter is the radio flux normalized by the optical flux. The dispersion in the relationship could be due to the differences in intrinsic AGN parameters, e.g., black hole mass, accretion process, etc. Such a strong positive correlation is found to be present in both high-z NLS1 ($r_s=0.800^{+0.003}_{-0.003}$) and low-z NLS1 samples ($r_s=0.727^{+0.005}_{-0.005}$), which agrees with the previous finding on low-z NLS1 sample \citep[e.g.,][]{2018MNRAS.480.1796S}. The radio-loudness of our high-$z$ 
sample ranges from $4-10^{4.1}$ with a median of 99, while for the low-$z$ sample the 
range is $1-10^{4.8}$ with a median of 17. The fraction of RL source is found to 
be similar (4.2\%) in the high-$z$ NLS1 sample and the low-$z$ NLS1 (2.8\%). The 
radio-luminosities of our high-$z$ sample range from $10^{31.1}-10^{34.7}$ 
erg s$^{-1}$ Hz$^{-1}$ with a median of $10^{32.4}$ erg s$^{-1}$ Hz$^{-1}$, much 
larger than the low-z sample, that ranges from $10^{29.1}-10^{34.1}$ erg s$^{-1}$ 
Hz$^{-1}$ with a median of $10^{30.9}$ erg s$^{-1}$ Hz$^{-1}$. The low-$z$ sample, which is mostly 
radio-quiet also has much lower radio-power and belongs to FR I, while high-$z$ 
NLS1s are mostly radio-loud with powerful radio jets and belongs to the 
FR II category.

\begin{figure}
                    \resizebox{9cm}{8cm}{\includegraphics{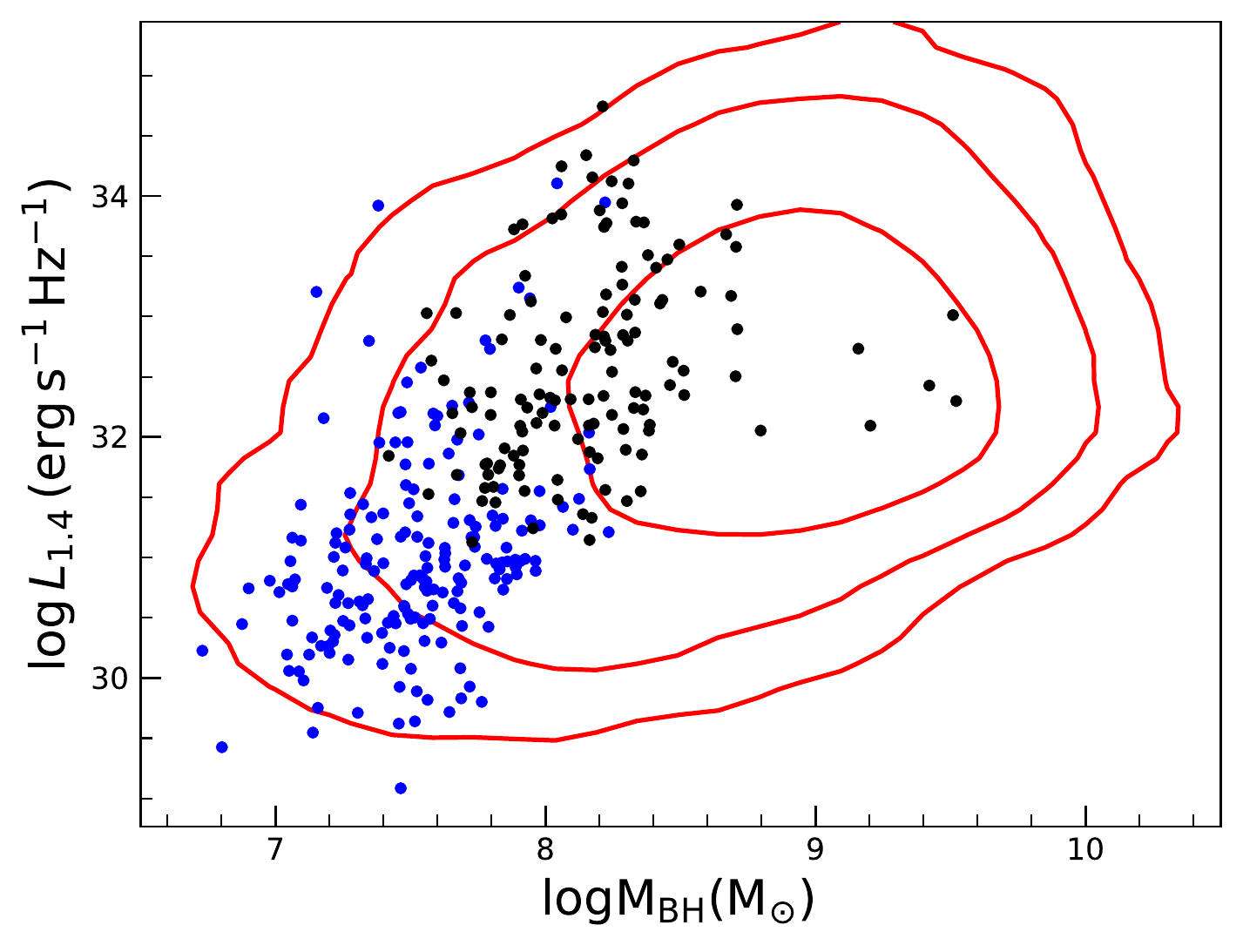}}  
                    \caption{The 1.4GHz radio luminosity is plotted against 
black hole mass from low-z (blue filled circles), high-z (black filled circles) 
and SDSS DR14 quasars for $z<2.2$ (as contours in red).}\label{Fig:L_RADIO_MBH} 
                  \end{figure}

We do not find any evidence for bimodality both in the NLS1 sample as well as 
in the parent sample of SDSS quasars. This is in contrast to the earlier reports on 
radio-loud and radio-quiet dichotomy \citep{1989AJ.....98.1195K}, however, our 
results agree with the previous findings which are based on a large sample 
\citep{Rafter_2008,2018BSRSL..87..379R,2018MNRAS.480.1796S}. The presence of 
bimodality in the earlier studies could be due to sample selection biases e.g., 
host galaxy type such as the radio-loud sources being hosted in giant elliptical 
and radio-quiet sources being hosted in spiral galaxies \citep{Sikora_2007}. Also, 
such bimodality may not be present if the nuclear radio emission is smoothly 
correlated with the AGN properties, such as black hole mass and Eddington ratio, 
as already found by \citet{Laor_2000} and \citet{Lacy_2001}. Indeed we found the 
radio luminosity to be strongly correlated with black hole mass (see Figure 
\ref{Fig:L_RADIO_MBH}) having $r_s=0.596^{+0.023}_{-0.022}$ for the NLS1s (including low-$z$ and 
high-$z$). This correlation is also present separately in the low-$z$ ($r_s=0.313^{+0.042}_{-0.043}$) 
and high-$z$ NLS1 samples ($r_s=0.394^{+0.048}_{-0.049}$) though a little weaker due to the sample 
range since high-$z$ radio detected NLS1s have in general black holes mass of 
about $10^8 M_{\odot}$. Moreover, the positive correlation between radio-power and black hole mass is 
also present albeit with a large dispersion ($r_s=0.218^{+0.003}_{-0.003}$) in the parent SDSS DR14 quasars 
sample supporting the finding of \citet{Laor_2000}. \citet{Lacy_2001} found this 
correlation to be tighter when the Eddington ratio is also considered along with 
black hole masses, however, the dependency is weaker at a low Eddington ratio. Since 
our NLS1s are drawn from the parent SDSS DR14 quasars sample and have a higher Eddington ratio, a tighter correlation in the former compared to the latter is thus expected.

       \begin{figure}
      \resizebox{9cm}{8cm}{\includegraphics{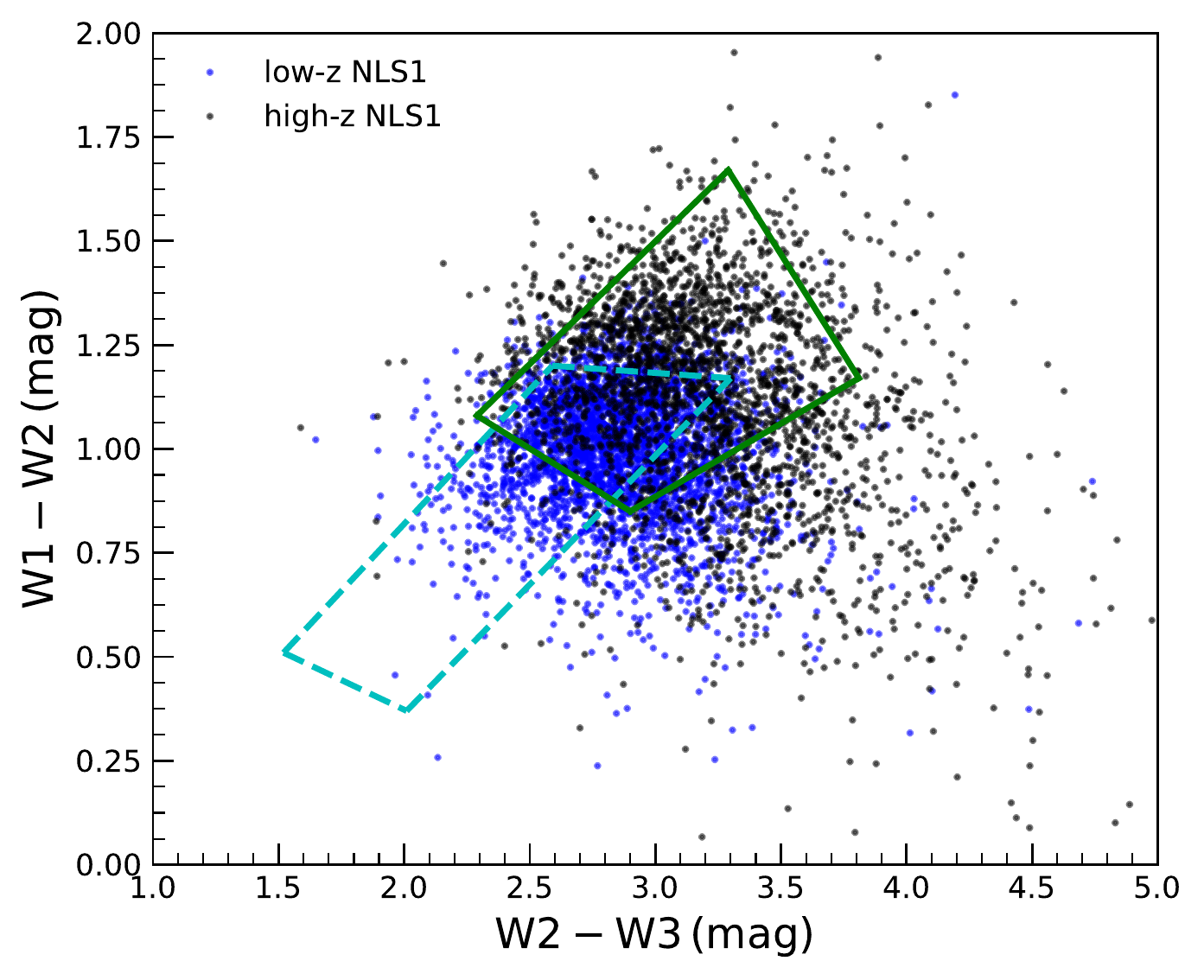}}  
      \caption{WISE W1-W2 color vs W2-W3 color diagram for NLS1s. Filled
blue and black circles are the low-$z$ and high-$z$ NLS1s respectively.
The location of the WISE $\gamma$-ray strips for FSRQ (solid green) and BL Lacs 
(dashed-cyan) are also shown.}\label{Fig:color-color} 
    \end{figure}
    
\subsection{color-color diagram}
In \citet{2019MNRAS.483.2362R}, we studied the infrared color and variability 
of low-$z$ NLS1s from SDSS DR12 using photometry magnitude from WISE. 
We found 57.6\% of the variable NLS1s are situated in the WISE color-color diagram, 
while 30\% are within the WISE $\gamma$-ray strip of BL Lac and FSRQs. To study the 
IR properties, we used the compilation of \citet{2018A&A...613A..51P} and 
\citetalias{2020ApJS..249...17R}, where SDSS DR14 quasars were cross-matched with the 
AllWISE Source catalog \citep{Wright_2010} with a matching radius of 
2$^{\prime\prime}$. About 92\% and 98\% of our high-$z$ and low-$z$ NLS1s, 
respectively, have WISE counterpart. In Figure \ref{Fig:color-color}, we plot 
the WISE W1-W2 color against W2-W3 color for the low-$z$ NLS1 (blue) and 
high-$z$ NLS1 (black) candidates. The mean W1-W2 color of high-$z$ (low-$z$) 
NLS1s is found to be $1.14\pm0.26$ ($1.00\pm0.16$), while W2-W3 color is 
$3.18\pm0.44$ ($2.86\pm0.31$). 

About 77\% of low-$z$ and 61\% of high-$z$ BLS1s  
lie in the region occupied by WISE $\gamma$-ray strip. About 60\% (16\%) of high-$z$ 
NLS1s are located in the region occupied by FSRQs (BL Lacs), among them, 15\% are 
common to BL Lac and FSRQs, while the same for the low-$z$ NLS1 sample is found 
to be 65\% (57\%) among which 45\% are common to both. Thus, based on the IR 
color-color diagram, it is evident that most of our NLS1 galaxies have blazar like 
properties and are AGN as expected since they are optically selected quasars 
with broad emission lines.

\begin{figure}
 \resizebox{9cm}{8cm}{\includegraphics{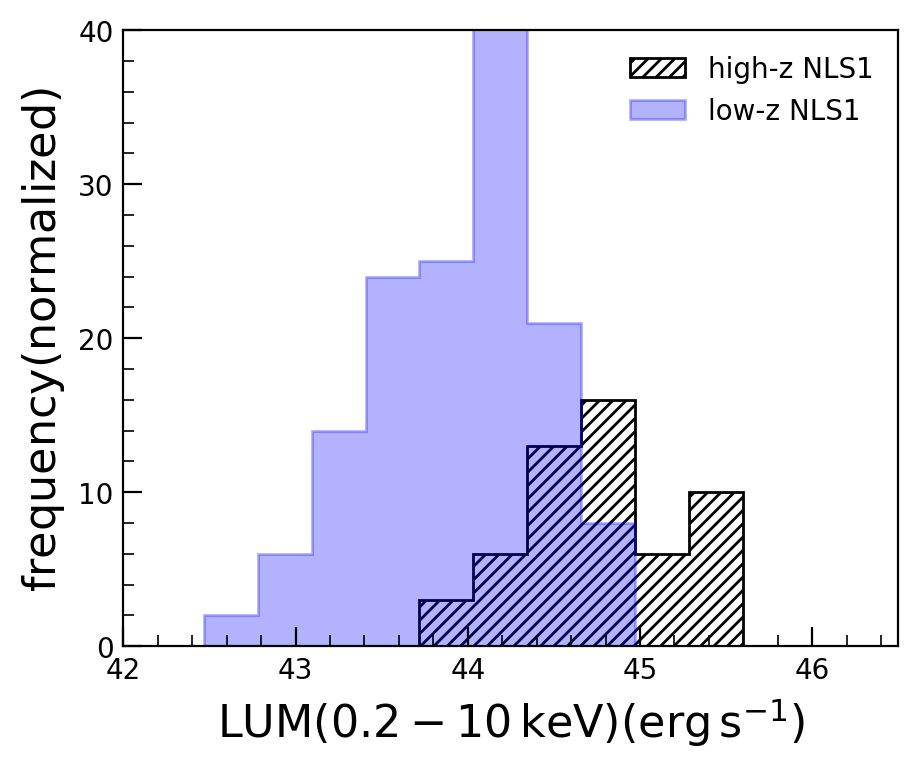}}  
 \caption{The soft X-ray 0.2-10 keV luminosity distribution of 
low-$z$ and high-$z$ NLS1s. }\label{Fig:LUM_X-ray} 
\end{figure}  
  
\subsection{X-ray properties}
 The steep soft X-ray spectra and soft X-ray access are the main characteristics 
of NLS1 galaxies. Recently, \citet{2020ApJ...896...95O} performed soft and hard X-ray 
spectral analysis of a large sample of NLS1 and BLS1 using the low-$z$ sample of 
\citet{2017ApJS..229...39R}. They found that the X-ray photon index strongly 
correlates with the Eddington ratio and R$_{\mathrm{Fe II}}$. It is also found to 
be anti-correlated with line width. \citetalias{2020ApJS..249...17R} culled multi-band 
properties of SDSS DR14 quasars from \citet{2018A&A...613A..51P} who cross-matched 
SDSS DR14 quasars with the Third XMM-Newton Serendipitous Source 
Catalog \citep[3XMM-DR7;][]{2016A&A...590A...1R} using a standard 
5.0$^{\prime\prime}$ matching radius.

To investigate the X-ray properties of high-$z$ NLS1, we used these values. Among 
the low-$z$ NLS1 sources having $z <0.8$, 141 are present in 3XMM-DR7, while only 
54 of the high-z NLS1 candidates are present. In Figure \ref{Fig:LUM_X-ray}, we 
plot the soft X-ray (0.2-10 keV) luminosity of the sample. The median 
$\log L_{0.2-10}$ keV of high-$z$ (low-$z$) NLS1 candidate is found to be 
$44.78\pm0.43$ ($44.03\pm0.49$) erg s$^{-1}$.

\begin{figure}
\resizebox{9cm}{5cm}{\includegraphics{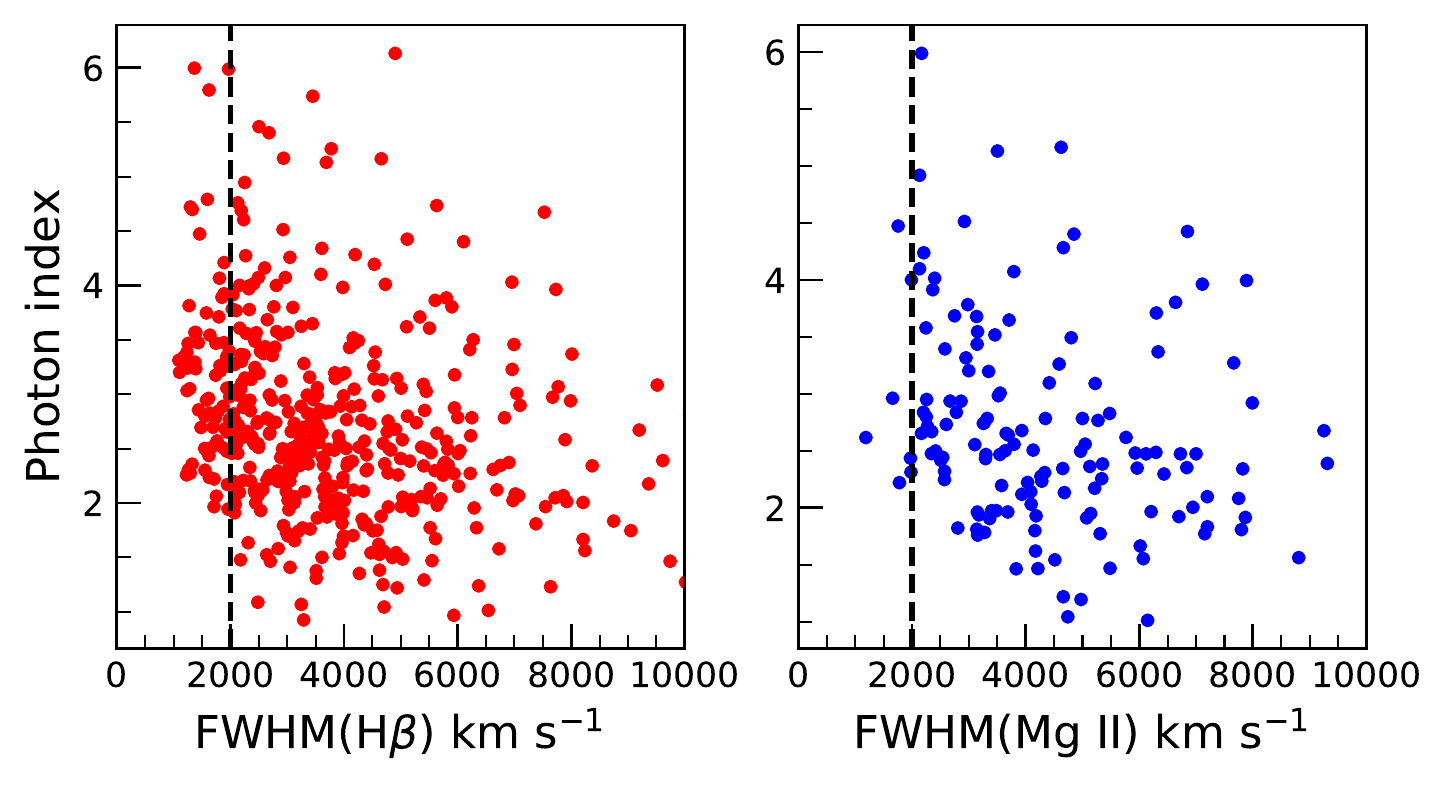}}  
\caption{The ROSAT soft X-ray (0.1-2 KeV) photon index is plotted against FWHM (H$\beta$) and FWHM(Mg II). The vertical line at FWHM=2000 km s$^{-1}$ represents the division of NLS1 from BLS1.}\label{Fig:photon_index} 
\end{figure}

We also cross-correlated the DR14 quasar spectral catalog of \citetalias{2020ApJS..249...17R} with the ROSAT all-sky survey (2RXS) source catalog \citep{2016A&A...588A.103B} within a search radius of 30$^{\prime\prime}$ and found 8,658 detections. The 2RXS catalog provides photon index from power-law spectral fitting that can be used to study the correlation of photo index with the emission line width. In Figure \ref{Fig:photon_index}, we plot soft X-ray (0.1-2 KeV) ROSAT photon index as a function of FWHM of H$\beta$ (left) and Mg II (right) for 459 and 143 sources, respectively, that have at least 50 source counts and photon index measurements better than 1$\sigma$. Note that the average uncertainty of ROSAT photon index is 50\%. We find a general trend of decreasing photon index with increasing line width, both for H$\beta$ and Mg II line with $r_s=-0.30$ and $r_s=-0.31$\footnote{We did not perform Monte Carlo due to large uncertainties in ROSAT photon indices.}, respectively. This is consistent with the fact of low-z NLS1s having a higher photon index compared to the BLS1s found in previous studies \citep[e.g.,][]{1996A&A...305...53B,2020ApJ...896...95O}. The anti-correlation of photon index with FWHM(Mg II) suggests that the latter is a good proxy for FWHM(H$\beta$) for selecting high-z NLS1s.

\section{Summary}\label{sec:summary}

Our knowledge of the general physical properties of NLS1s are based
on NLS1s available up to $z$ = 0.8. This is mainly constrained by the definition  of
NLS1s that requires the presence of  H$\beta$ emission line, which is only 
available in large optical spectroscopic surveys up to sources with 
$z < 0.8$. To find out high-$z$ NLS1 candidates, we studied H$\beta$ and Mg II 
line properties of a large number of SDSS DR14 quasars using the spectral 
properties cataloged by \citetalias{2020ApJS..249...17R}. We found that the UV EV1 has a similar shape as the optical plane of EV1 which are believed to be the main 
driver of the quasars spectral properties. The shape of UV EV1 is due to the 
strong correlation between the EW and FWHM of Mg II line while the shape of optical 
EV1 is due to the strong anti-correlation between the EW of Fe II in the optical 
against the FWMH of H$\beta$. A strong correlation between Mg II and H$\beta$ 
line FWHM is 
also found which allowed us to provide a sample of high-$z$ NLS1 based on Mg II 
FWHM $<2000$ km s$^{-1}$.

We supplemented this work with a low-$z$ NLS1 sample 
based on the H$\beta$ line as per the classical definition. The high-$z$ NLS1 sample has a redshift range of $0.8-2.5$ and median logarithmic bolometric luminosity of 
$46.16\pm0.42$ erg s$^{-1}$ larger than the median logarithmic bolometric luminosity of $45.29\pm0.39$ erg s$^{-1}$ found for the low-z sample. The median logarithmic black hole mass of the 
sample is $8.01\pm0.35$ $M_{\odot}$ and logarithmic Eddington ratio is $0.02\pm0.27$. The 
black hole mass and Eddington ratio of the high-$z$ NLS1 is slightly higher and lower, respectively, compared to the low-$z$ 
sample due to the higher median bolometric luminosity in the former but compared to the parent SDSS Dr14 quasar sample, the black hole mass is much 
lower and the Eddington ratio is much higher.

The radio detection rate of high-$z$ 
and low-$z$ samples are similar to the SDSS DR14 quasars at a similar redshift range. 
The high-$z$ NLS1 sources are mainly FR II type sources having powerful jets 
compared to their low-z NLS1 counterparts which seem to be due to their higher 
black hole mass since a strong correlation between 1.4GHz radio luminosity and 
black hole mass has been found. A major fraction of them is located within the 
WISE $\gamma$-ray strip similar to the low-$z$ NLS1s confirming their AGN nature. 
This catalog will be immensely useful to the community in studies of high-z NLS1s e.g., comparison with BLS1s, and the search for high-z $\gamma$-ray emitters.

\acknowledgements
We thank the referee for valuable comments/suggestions that helped to improve the quality of our paper. JK acknowledges financial support from the Academy of Finland, grant 311438. JS was supported by Basic Science Research Program through the National Research Foundation of Korea (NRF) funded by the Ministry of Education (2019R1A6A3A01093189). SR thanks Neha Sharma (FINCA) for carefully reading the manuscript.

     \bibliographystyle{apj}
     \bibliography{ref}

\begin{thebibliography}{}
\expandafter\ifx\csname natexlab\endcsname\relax\def\natexlab#1{#1}\fi

\bibitem[{{Akritas} \& {Bershady}(1996)}]{1996ApJ...470..706A}
{Akritas}, M.~G., \& {Bershady}, M.~A. 1996, \apj, 470, 706

\bibitem[{{Baldi} {et~al.}(2016){Baldi}, {Capetti}, {Robinson}, {Laor}, \&
  {Behar}}]{2016MNRAS.458L..69B}
{Baldi}, R.~D., {Capetti}, A., {Robinson}, A., {Laor}, A., \& {Behar}, E. 2016,
  \mnras, 458, L69

\bibitem[{{Becker} {et~al.}(1995){Becker}, {White}, \&
  {Helfand}}]{1995ApJ...450..559B}
{Becker}, R.~H., {White}, R.~L., \& {Helfand}, D.~J. 1995, \apj, 450, 559

\bibitem[{{Boller} {et~al.}(1996){Boller}, {Brandt}, \&
  {Fink}}]{1996A&A...305...53B}
{Boller}, T., {Brandt}, W.~N., \& {Fink}, H. 1996, \aap, 305, 53

\bibitem[{{Boller} {et~al.}(2016){Boller}, {Freyberg}, {Tr{\"u}mper}, {Haberl},
  {Voges}, \& {Nandra}}]{2016A&A...588A.103B}
{Boller}, T., {Freyberg}, M.~J., {Tr{\"u}mper}, J., {et~al.} 2016, \aap, 588,
  A103

\bibitem[{{Boroson} \& {Green}(1992)}]{1992ApJS...80..109B}
{Boroson}, T.~A., \& {Green}, R.~F. 1992, \apjs, 80, 109

\bibitem[{{Calderone} {et~al.}(2013){Calderone}, {Ghisellini}, {Colpi}, \&
  {Dotti}}]{2013MNRAS.431..210C}
{Calderone}, G., {Ghisellini}, G., {Colpi}, M., \& {Dotti}, M. 2013, \mnras,
  431, 210

\bibitem[{{Chen} {et~al.}(2018){Chen}, {Berton}, {La Mura}, {Congiu}, {Cracco},
  {Foschini}, {Fan}, {Ciroi}, {Rafanelli}, \& {Bastieri}}]{2018A&A...615A.167C}
{Chen}, S., {Berton}, M., {La Mura}, G., {et~al.} 2018, \aap, 615, A167

\bibitem[{{Curran}(2014)}]{2014arXiv1411.3816C}
{Curran}, P.~A. 2014, arXiv e-prints, arXiv:1411.3816

\bibitem[{{Decarli} {et~al.}(2008){Decarli}, {Dotti}, {Fontana}, \&
  {Haardt}}]{2008MNRAS.386L..15D}
{Decarli}, R., {Dotti}, M., {Fontana}, M., \& {Haardt}, F. 2008, \mnras, 386,
  L15

\bibitem[{{Denney}(2012)}]{2012ApJ...759...44D}
{Denney}, K.~D. 2012, \apj, 759, 44

\bibitem[{{Dong} {et~al.}(2009){Dong}, {Wang}, {Wang}, {Fan}, {Wang}, {Zhou},
  \& {Yuan}}]{2009ApJ...703L...1D}
{Dong}, X.-B., {Wang}, T.-G., {Wang}, J.-G., {et~al.} 2009, \apjl, 703, L1

\bibitem[{{Donoso} {et~al.}(2009){Donoso}, {Best}, \&
  {Kauffmann}}]{2009MNRAS.392..617D}
{Donoso}, E., {Best}, P.~N., \& {Kauffmann}, G. 2009, \mnras, 392, 617

\bibitem[{{Fanaroff} \& {Riley}(1974)}]{1974MNRAS.167P..31F}
{Fanaroff}, B.~L., \& {Riley}, J.~M. 1974, \mnras, 167, 31P

\bibitem[{Ferland {et~al.}(2009)Ferland, Hu, Wang, Baldwin, Porter, van Hoof,
  \& Williams}]{Ferland_2009}
Ferland, G.~J., Hu, C., Wang, J.-M., {et~al.} 2009, The Astrophysical Journal,
  707, L82

\bibitem[{{Goodrich}(1989)}]{1989ApJ...342..224G}
{Goodrich}, R.~W. 1989, \apj, 342, 224

\bibitem[{{Grupe}(2000)}]{2000NewAR..44..455G}
{Grupe}, D. 2000, \nar, 44, 455

\bibitem[{{Grupe} \& {Mathur}(2004)}]{2004ApJ...606L..41G}
{Grupe}, D., \& {Mathur}, S. 2004, \apjl, 606, L41

\bibitem[{{Guo} {et~al.}(2018){Guo}, {Shen}, \& {Wang}}]{2018ascl.soft09008G}
{Guo}, H., {Shen}, Y., \& {Wang}, S. 2018, {PyQSOFit: Python code to fit the
  spectrum of quasars}, Astrophysics Source Code Library, ascl:1809.008

\bibitem[{{Halpern} {et~al.}(1998){Halpern}, {Eracleous}, \&
  {Forster}}]{1998ApJ...501..103H}
{Halpern}, J.~P., {Eracleous}, M., \& {Forster}, K. 1998, \apj, 501, 103

\bibitem[{{Halpern} \& {Moran}(1998)}]{1998ApJ...494..194H}
{Halpern}, J.~P., \& {Moran}, E.~C. 1998, \apj, 494, 194

\bibitem[{{Halpern} {et~al.}(1999){Halpern}, {Turner}, \&
  {George}}]{1999MNRAS.307L..47H}
{Halpern}, J.~P., {Turner}, T.~J., \& {George}, I.~M. 1999, \mnras, 307, L47

\bibitem[{{J{\"a}rvel{\"a}} {et~al.}(2018){J{\"a}rvel{\"a}},
  {L{\"a}hteenm{\"a}ki}, \& {Berton}}]{2018A&A...619A..69J}
{J{\"a}rvel{\"a}}, E., {L{\"a}hteenm{\"a}ki}, A., \& {Berton}, M. 2018, \aap,
  619, A69

\bibitem[{{Joly}(1987)}]{1987A&A...184...33J}
{Joly}, M. 1987, \aap, 184, 33

\bibitem[{{Jun} {et~al.}(2015){Jun}, {Im}, {Lee}, {Ohyama}, {Woo}, {Fan},
  {Goto}, {Kim}, {Kim}, {Kim}, {Lee}, {Nakagawa}, {Pearson}, \&
  {Serjeant}}]{2015ApJ...806..109J}
{Jun}, H.~D., {Im}, M., {Lee}, H.~M., {et~al.} 2015, \apj, 806, 109

\bibitem[{{Kellermann} {et~al.}(1989){Kellermann}, {Sramek}, {Schmidt},
  {Shaffer}, \& {Green}}]{1989AJ.....98.1195K}
{Kellermann}, K.~I., {Sramek}, R., {Schmidt}, M., {Shaffer}, D.~B., \& {Green},
  R. 1989, \aj, 98, 1195

\bibitem[{{Kelly}(2007)}]{2007ApJ...665.1489K}
{Kelly}, B.~C. 2007, \apj, 665, 1489

\bibitem[{{Komossa} {et~al.}(2006){Komossa}, {Voges}, {Xu}, {Mathur}, {Adorf},
  {Lemson}, {Duschl}, \& {Grupe}}]{2006AJ....132..531K}
{Komossa}, S., {Voges}, W., {Xu}, D., {et~al.} 2006, \aj, 132, 531

\bibitem[{{Kova{\v c}evi{\'c}-Doj{\v c}inovi{\'c}} \&
  {Popovi{\'c}}(2015)}]{2015ApJS..221...35K}
{Kova{\v c}evi{\'c}-Doj{\v c}inovi{\'c}}, J., \& {Popovi{\'c}}, L.~{\v C}.
  2015, \apjs, 221, 35

\bibitem[{Lacy {et~al.}(2001)Lacy, Laurent-Muehleisen, Ridgway, Becker, \&
  White}]{Lacy_2001}
Lacy, M., Laurent-Muehleisen, S.~A., Ridgway, S.~E., Becker, R.~H., \& White,
  R.~L. 2001, The Astrophysical Journal, 551, L17

\bibitem[{Laor(2000)}]{Laor_2000}
Laor, A. 2000, The Astrophysical Journal, 543, L111

\bibitem[{{Leighly}(1999)}]{1999ApJS..125..317L}
{Leighly}, K.~M. 1999, \apjs, 125, 317

\bibitem[{{Marziani} {et~al.}(2018){Marziani}, {Dultzin}, {Sulentic}, {Del
  Olmo}, {Negrete}, {Mart{\'\i}nez-Aldama}, {D'Onofrio}, {Bon}, {Bon}, \&
  {Stirpe}}]{2018FrASS...5....6M}
{Marziani}, P., {Dultzin}, D., {Sulentic}, J.~W., {et~al.} 2018, Frontiers in
  Astronomy and Space Sciences, 5, 6

\bibitem[{{Mathur}(2000)}]{2000MNRAS.314L..17M}
{Mathur}, S. 2000, \mnras, 314, L17

\bibitem[{{Nandra} \& {Pounds}(1994)}]{1994MNRAS.268..405N}
{Nandra}, K., \& {Pounds}, K.~A. 1994, \mnras, 268, 405

\bibitem[{{Negrete} {et~al.}(2018){Negrete}, {Dultzin}, {Marziani}, {Esparza},
  {Sulentic}, {del Olmo}, {Mart{\'\i}nez-Aldama}, {Garc{\'\i}a L{\'o}pez},
  {D'Onofrio}, {Bon}, \& {Bon}}]{2018A&A...620A.118N}
{Negrete}, C.~A., {Dultzin}, D., {Marziani}, P., {et~al.} 2018, \aap, 620, A118

\bibitem[{{Nemmen} {et~al.}(2012){Nemmen}, {Georganopoulos}, {Guiriec},
  {Meyer}, {Gehrels}, \& {Sambruna}}]{2012Sci...338.1445N}
{Nemmen}, R.~S., {Georganopoulos}, M., {Guiriec}, S., {et~al.} 2012, Science,
  338, 1445

\bibitem[{{Netzer} {et~al.}(2004){Netzer}, {Shemmer}, {Maiolino}, {Oliva},
  {Croom}, {Corbett}, \& {di Fabrizio}}]{2004ApJ...614..558N}
{Netzer}, H., {Shemmer}, O., {Maiolino}, R., {et~al.} 2004, \apj, 614, 558

\bibitem[{{Netzer} \& {Trakhtenbrot}(2007)}]{2007ApJ...654..754N}
{Netzer}, H., \& {Trakhtenbrot}, B. 2007, \apj, 654, 754

\bibitem[{{Ojha} {et~al.}(2020){Ojha}, {Chand}, {Dewangan}, \&
  {Rakshit}}]{2020ApJ...896...95O}
{Ojha}, V., {Chand}, H., {Dewangan}, G.~C., \& {Rakshit}, S. 2020, \apj, 896,
  95

\bibitem[{{Olgu{\'\i}n-Iglesias} {et~al.}(2020){Olgu{\'\i}n-Iglesias},
  {Kotilainen}, \& {Chavushyan}}]{2020MNRAS.492.1450O}
{Olgu{\'\i}n-Iglesias}, A., {Kotilainen}, J., \& {Chavushyan}, V. 2020, \mnras,
  492, 1450

\bibitem[{{Osterbrock} \& {Dahari}(1983)}]{1983ApJ...273..478O}
{Osterbrock}, D.~E., \& {Dahari}, O. 1983, \apj, 273, 478

\bibitem[{{Paliya} {et~al.}(2018){Paliya}, {Ajello}, {Rakshit}, {Mand al},
  {Stalin}, {Kaur}, \& {Hartmann}}]{2018ApJ...853L...2P}
{Paliya}, V.~S., {Ajello}, M., {Rakshit}, S., {et~al.} 2018, \apjl, 853, L2

\bibitem[{{Paliya} {et~al.}(2019){Paliya}, {Parker}, {Jiang}, {Fabian},
  {Brenneman}, {Ajello}, \& {Hartmann}}]{2019ApJ...872..169P}
{Paliya}, V.~S., {Parker}, M.~L., {Jiang}, J., {et~al.} 2019, \apj, 872, 169

\bibitem[{{Paliya} {et~al.}(2013){Paliya}, {Stalin}, {Shukla}, \&
  {Sahayanathan}}]{2013ApJ...768...52P}
{Paliya}, V.~S., {Stalin}, C.~S., {Shukla}, A., \& {Sahayanathan}, S. 2013,
  \apj, 768, 52

\bibitem[{{P{\^a}ris} {et~al.}(2018){P{\^a}ris}, {Petitjean}, {Aubourg},
  {Myers}, {Streblyanska}, {Lyke}, {Anderson}, {Armengaud}, {Bautista},
  {Blanton}, {Blomqvist}, {Brinkmann}, {Brownstein}, {Brand t}, {Burtin},
  {Dawson}, {de la Torre}, {Georgakakis}, {Gil-Mar{\'\i}n}, {Green}, {Hall},
  {Kneib}, {LaMassa}, {Le Goff}, {MacLeod}, {Mariappan}, {McGreer}, {Merloni},
  {Noterdaeme}, {Palanque-Delabrouille}, {Percival}, {Ross}, {Rossi},
  {Schneider}, {Seo}, {Tojeiro}, {Weaver}, {Weijmans}, {Y{\`e}che}, {Zarrouk},
  \& {Zhao}}]{2018A&A...613A..51P}
{P{\^a}ris}, I., {Petitjean}, P., {Aubourg}, {\'E}., {et~al.} 2018, \aap, 613,
  A51

\bibitem[{{Park} {et~al.}(2017){Park}, {Barth}, {Woo}, {Malkan}, {Treu},
  {Bennert}, {Assef}, \& {Pancoast}}]{2017ApJ...839...93P}
{Park}, D., {Barth}, A.~J., {Woo}, J.-H., {et~al.} 2017, \apj, 839, 93

\bibitem[{Rafter {et~al.}(2008)Rafter, Crenshaw, \& Wiita}]{Rafter_2008}
Rafter, S.~E., Crenshaw, D.~M., \& Wiita, P.~J. 2008, The Astronomical Journal,
  137, 42

\bibitem[{{Rakshit} {et~al.}(2019){Rakshit}, {Johnson}, {Stalin}, {Gand hi}, \&
  {Hoenig}}]{2019MNRAS.483.2362R}
{Rakshit}, S., {Johnson}, A., {Stalin}, C.~S., {Gand hi}, P., \& {Hoenig}, S.
  2019, \mnras, 483, 2362

\bibitem[{{Rakshit} \& {Stalin}(2017)}]{2017ApJ...842...96R}
{Rakshit}, S., \& {Stalin}, C.~S. 2017, \apj, 842, 96

\bibitem[{{Rakshit} {et~al.}(2017){Rakshit}, {Stalin}, {Chand}, \&
  {Zhang}}]{2017ApJS..229...39R}
{Rakshit}, S., {Stalin}, C.~S., {Chand}, H., \& {Zhang}, X.-G. 2017, \apjs,
  229, 39

\bibitem[{{Rakshit} {et~al.}(2018{\natexlab{a}}){Rakshit}, {Stalin}, {Chand},
  \& {Zhang}}]{2018BSRSL..87..379R}
---. 2018{\natexlab{a}}, Bulletin de la Societe Royale des Sciences de Liege,
  87, 379

\bibitem[{{Rakshit} {et~al.}(2018{\natexlab{b}}){Rakshit}, {Stalin}, {Hota}, \&
  {Konar}}]{2018ApJ...869..173R}
{Rakshit}, S., {Stalin}, C.~S., {Hota}, A., \& {Konar}, C. 2018{\natexlab{b}},
  \apj, 869, 173

\bibitem[{{Rakshit} {et~al.}(2020){Rakshit}, {Stalin}, \&
  {Kotilainen}}]{2020ApJS..249...17R}
{Rakshit}, S., {Stalin}, C.~S., \& {Kotilainen}, J. 2020, \apjs, 249, 17

\bibitem[{{Rosen} {et~al.}(2016){Rosen}, {Webb}, {Watson}, {Ballet}, {Barret},
  {Braito}, {Carrera}, {Ceballos}, {Coriat}, {Della Ceca}, {Denkinson},
  {Esquej}, {Farrell}, {Freyberg}, {Gris{\'e}}, {Guillout}, {Heil},
  {Koliopanos}, {Law-Green}, {Lamer}, {Lin}, {Martino}, {Michel}, {Motch},
  {Nebot Gomez-Moran}, {Page}, {Page}, {Page}, {Pakull}, {Pye}, {Read},
  {Rodriguez}, {Sakano}, {Saxton}, {Schwope}, {Scott}, {Sturm}, {Traulsen},
  {Yershov}, \& {Zolotukhin}}]{2016A&A...590A...1R}
{Rosen}, S.~R., {Webb}, N.~A., {Watson}, M.~G., {et~al.} 2016, \aap, 590, A1

\bibitem[{{Salviander} {et~al.}(2007){Salviander}, {Shields}, {Gebhardt}, \&
  {Bonning}}]{2007ApJ...662..131S}
{Salviander}, S., {Shields}, G.~A., {Gebhardt}, K., \& {Bonning}, E.~W. 2007,
  \apj, 662, 131

\bibitem[{{Sameshima} {et~al.}(2011){Sameshima}, {Kawara}, {Matsuoka}, {Oyabu},
  {Asami}, \& {Ienaka}}]{2011MNRAS.410.1018S}
{Sameshima}, H., {Kawara}, K., {Matsuoka}, Y., {et~al.} 2011, \mnras, 410, 1018

\bibitem[{{Sani} {et~al.}(2010){Sani}, {Lutz}, {Risaliti}, {Netzer}, {Gallo},
  {Trakhtenbrot}, {Sturm}, \& {Boller}}]{2010MNRAS.403.1246S}
{Sani}, E., {Lutz}, D., {Risaliti}, G., {et~al.} 2010, \mnras, 403, 1246

\bibitem[{{Shen} \& {Ho}(2014)}]{2014Natur.513..210S}
{Shen}, Y., \& {Ho}, L.~C. 2014, \nat, 513, 210

\bibitem[{{Shin} {et~al.}(2019){Shin}, {Nagao}, {Woo}, \&
  {Le}}]{2019ApJ...874...22S}
{Shin}, J., {Nagao}, T., {Woo}, J.-H., \& {Le}, H. A.~N. 2019, \apj, 874, 22

\bibitem[{{Sikora} {et~al.}(2007){Sikora}, {Stawarz}, \&
  {Lasota}}]{2007ApJ...658..815S}
{Sikora}, M., {Stawarz}, {\L}., \& {Lasota}, J.-P. 2007, \apj, 658, 815

\bibitem[{Sikora {et~al.}(2007)Sikora, Stawarz, \& Lasota}]{Sikora_2007}
Sikora, M., Stawarz, {\L}., \& Lasota, J.-P. 2007, The Astrophysical Journal,
  658, 815

\bibitem[{{Singh} \& {Chand}(2018)}]{2018MNRAS.480.1796S}
{Singh}, V., \& {Chand}, H. 2018, \mnras, 480, 1796

\bibitem[{{Sulentic} {et~al.}(2000){Sulentic}, {Marziani}, \&
  {Dultzin-Hacyan}}]{2000ARA&A..38..521S}
{Sulentic}, J.~W., {Marziani}, P., \& {Dultzin-Hacyan}, D. 2000, \araa, 38, 521

\bibitem[{{Trakhtenbrot} \& {Netzer}(2012)}]{2012MNRAS.427.3081T}
{Trakhtenbrot}, B., \& {Netzer}, H. 2012, \mnras, 427, 3081

\bibitem[{{Tsuzuki} {et~al.}(2006){Tsuzuki}, {Kawara}, {Yoshii}, {Oyabu},
  {Tanab{\'e}}, \& {Matsuoka}}]{2006ApJ...650...57T}
{Tsuzuki}, Y., {Kawara}, K., {Yoshii}, Y., {et~al.} 2006, \apj, 650, 57

\bibitem[{{V{\'e}ron-Cetty} {et~al.}(2001){V{\'e}ron-Cetty}, {V{\'e}ron}, \&
  {Gon{\c c}alves}}]{2001A&A...372..730V}
{V{\'e}ron-Cetty}, M.-P., {V{\'e}ron}, P., \& {Gon{\c c}alves}, A.~C. 2001,
  \aap, 372, 730

\bibitem[{{Vestergaard} \& {Wilkes}(2001)}]{2001ApJS..134....1V}
{Vestergaard}, M., \& {Wilkes}, B.~J. 2001, \apjs, 134, 1

\bibitem[{{Viswanath} {et~al.}(2019){Viswanath}, {Stalin}, {Rakshit}, {Kurian},
  {Ujjwal}, {Gudennavar}, \& {Kartha}}]{2019ApJ...881L..24V}
{Viswanath}, G., {Stalin}, C.~S., {Rakshit}, S., {et~al.} 2019, \apjl, 881, L24

\bibitem[{{Williams} {et~al.}(2002){Williams}, {Pogge}, \&
  {Mathur}}]{2002AJ....124.3042W}
{Williams}, R.~J., {Pogge}, R.~W., \& {Mathur}, S. 2002, \aj, 124, 3042

\bibitem[{Wright {et~al.}(2010)Wright, Eisenhardt, Mainzer, Ressler, Cutri,
  Jarrett, Kirkpatrick, Padgett, McMillan, Skrutskie, Stanford, Cohen, Walker,
  Mather, Leisawitz, Gautier, McLean, Benford, Lonsdale, Blain, Mendez, Irace,
  Duval, Liu, Royer, Heinrichsen, Howard, Shannon, Kendall, Walsh, Larsen,
  Cardon, Schick, Schwalm, Abid, Fabinsky, Naes, \& Tsai}]{Wright_2010}
Wright, E.~L., Eisenhardt, P. R.~M., Mainzer, A.~K., {et~al.} 2010, The
  Astronomical Journal, 140, 1868

\bibitem[{{Xu} {et~al.}(2012){Xu}, {Komossa}, {Zhou}, {Lu}, {Li}, {Grupe},
  {Wang}, \& {Yuan}}]{2012AJ....143...83X}
{Xu}, D., {Komossa}, S., {Zhou}, H., {et~al.} 2012, \aj, 143, 83

\bibitem[{{Yao} {et~al.}(2019){Yao}, {Komossa}, {Liu}, {Yi}, {Yuan}, {Zhou}, \&
  {Wu}}]{2019MNRAS.487L..40Y}
{Yao}, S., {Komossa}, S., {Liu}, W.-J., {et~al.} 2019, \mnras, 487, L40

\bibitem[{{Yao} {et~al.}(2015){Yao}, {Yuan}, {Zhou}, {Komossa}, {Zhang},
  {Qiao}, \& {Liu}}]{2015MNRAS.454L..16Y}
{Yao}, S., {Yuan}, W., {Zhou}, H., {et~al.} 2015, \mnras, 454, L16

\bibitem[{{Yip} {et~al.}(2004){Yip}, {Connolly}, {Vanden Berk}, {Ma},
  {Frieman}, {SubbaRao}, {Szalay}, {Richards}, {Hall}, {Schneider}, {Hopkins},
  {Trump}, \& {Brinkmann}}]{2004AJ....128.2603Y}
{Yip}, C.~W., {Connolly}, A.~J., {Vanden Berk}, D.~E., {et~al.} 2004, \aj, 128,
  2603

\bibitem[{{Zhou} {et~al.}(2006){Zhou}, {Wang}, {Yuan}, {Lu}, {Dong}, {Wang}, \&
  {Lu}}]{2006ApJS..166..128Z}
{Zhou}, H., {Wang}, T., {Yuan}, W., {et~al.} 2006, \apjs, 166, 128

\end{thebibliography}

\end{document}